\documentclass[acmsmall,screen]{acmart}
\usepackage{enumitem}
\usepackage{hyperref}
\usepackage{multirow}
\usepackage{amsthm}
\usepackage{array}
\usepackage{subfigure}
\usepackage{graphicx}

\usepackage{pifont}

\setlist[enumerate]{align=left}

\usepackage{algorithm}
\usepackage{algorithmic}

\newtheorem{innercustomgeneric}{\customgenericname}
\providecommand{\customgenericname}{}
\newcommand{\newcustomtheorem}[2]{%
  \newenvironment{#1}[1]
  {%
  \renewcommand\customgenericname{#2}%
  \renewcommand\theinnercustomgeneric{##1}%
  \innercustomgeneric
  }
  {\endinnercustomgeneric}
}
\newcustomtheorem{myprop}{Proposition}
\newcustomtheorem{mytheo}{Theorem}
\newcustomtheorem{mydef}{Definition}

\newcommand{\tabincell}[2]{\begin{tabular}{@{}#1@{}}#2\end{tabular}}  
\AtBeginDocument{%
  \providecommand\BibTeX{{%
    \normalfont B\kern-0.5em{\scshape i\kern-0.25em b}\kern-0.8em\TeX}}}

\setcopyright{acmcopyright}
\copyrightyear{2023}
\acmYear{2023}
\acmDOI{10.1145/0000001.0000001}

\acmJournal{{TOIS}}
\acmVolume{99}
\acmNumber{99}
\acmArticle{99}
\acmMonth{1}

\begin{document}

\title{Bounding System-Induced Biases in Recommender Systems with A Randomized Dataset}

\author{Dugang Liu}
\additionalaffiliation{%
  \institution{Guangdong Laboratory of Artificial Intelligence and Digital Economy (SZ)}
}
\affiliation{%
  \institution{College of Computer Science and Software Engineering, Shenzhen University}
  \streetaddress{3688\# Nanhai Avenue}
  \city{Shenzhen}                                                    
  \state{Guangdong}
  \country{China}
  \postcode{518060}}
\email{dugang.ldg@gmail.com}
\author{Pengxiang Cheng}
\affiliation{%
  \institution{Huawei Noah's Ark Lab}
  \streetaddress{Bantian Street}
  \city{Shenzhen}
  \state{Guangdong}
  \country{China}
  \postcode{518129}}
\email{pengxiang.cpx@gmail.com}
\author{Zinan Lin}
\affiliation{%
  \institution{College of Computer Science and Software Engineering, Shenzhen University}
  \streetaddress{3688\# Nanhai Avenue}
  \city{Shenzhen}                                                    
  \state{Guangdong}
  \country{China}
  \postcode{518060}}
\email{lzn87591@gmail.com}
\author{Xiaolian Zhang}
\author{Zhenhua Dong}
\affiliation{%
  \institution{Huawei 2012 Lab}
  \streetaddress{Bantian Street}
  \city{Shenzhen}
  \state{Guangdong}
  \country{China}
  \postcode{518129}}
\email{zhangxiaolian@huawei.com}
\email{dongzhenhua@huawei.com}
\author{Rui Zhang}
\affiliation{%
  \institution{Tsinghua University}
  \state{Guangdong}
  \country{China}
  \postcode{518055}}
\email{rayteam@yeah.net}
\email{https://ruizhang.info/}
\author{Xiuqiang He}
\affiliation{%
  \institution{Tencent FIT}
  \streetaddress{33\# Haitian Second Road}
  \city{Shenzhen}
  \state{Guangdong}
  \country{China}
  \postcode{518057}}
\email{xiuqianghe@tencent.com}
\author{Weike Pan}
\authornote{Co-corresponding authors.}
\author{Zhong Ming}
\authornotemark[2]
\affiliation{%
  \institution{College of Computer Science and Software Engineering, Shenzhen University}
  \streetaddress{3688\# Nanhai Avenue}
  \city{Shenzhen}
  \state{Guangdong}
  \country{China}
  \postcode{518060}}
\email{panweike@szu.edu.cn}
\email{mingz@szu.edu.cn}

\renewcommand{\shortauthors}{Liu, et al.}

\begin{abstract}
  Debiased recommendation with a randomized dataset has shown very promising results in mitigating the system-induced biases. However, it still lacks more theoretical insights or an ideal optimization objective function compared with the other more well studied route without a randomized dataset. To bridge this gap, we study the debiasing problem from a new perspective and propose to directly minimize the upper bound of an ideal objective function, which facilitates a better potential solution to the system-induced biases. Firstly, we formulate a new ideal optimization objective function with a randomized dataset. Secondly, according to the prior constraints that an adopted loss function may satisfy, we derive two different upper bounds of the objective function, i.e., a generalization error bound with the triangle inequality and a generalization error bound with the separability. Thirdly, we show that most existing related methods can be regarded as the insufficient optimization of these two upper bounds. Fourthly, we propose a novel method called debiasing approximate upper bound with a randomized dataset (DUB), which achieves a more sufficient optimization of these upper bounds. Finally, we conduct extensive experiments on a public dataset and a real product dataset to verify the effectiveness of our DUB.
\end{abstract}



\maketitle

\section{Introduction}\label{intro}
\begin{figure}[htbp]
\centering
\includegraphics[width=1.0\columnwidth]{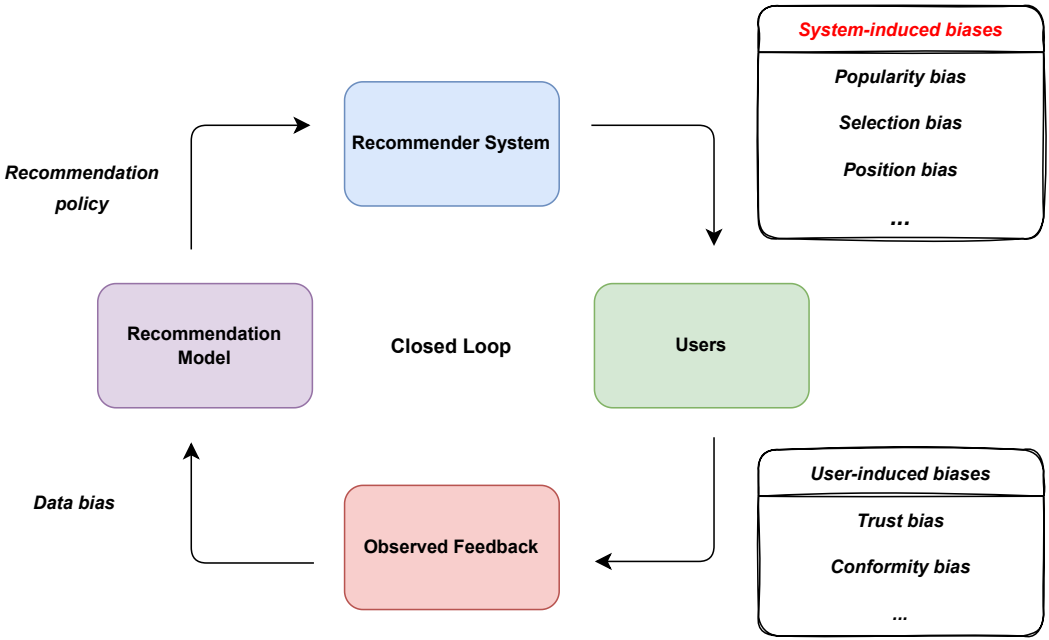}
\caption{The feedback loop in a recommender system, where the observed feedback contains the data bias coupled by the system-induced biases and the user-induced biases. The former is caused by the stochastic recommendation policy deployed on a recommender system, and the latter depends on the user characteristics.}
\label{fig:loop}
\end{figure}

Recently, the bias issue in recommender systems has received more attention from both of the research communities and industries~\cite{zhang2020large,morik2020controlling,shivaswamy2021bias,wu2021unbiased,yadav2021policy,oosterhuis2021unifying}. 
Intuitively, as shown in Figure~\ref{fig:loop}, a user will experience system-induced biases and user-induced biases when interacting with a recommender system. 
The \textit{system-induced biases} are caused by the stochastic recommendation policy deployed on a recommender system, and the selection and display order of each item is treated differently by the policy, including popularity bias~\cite{abdollahpouri2017controlling,canamares2018should,zhu2021popularity}, selection bias~\cite{schnabel2016recommendations,ovaisi2020correcting,lee2021dual} and position bias~\cite{wang2018position,agarwal2019estimating}, etc.
The \textit{user-induced biases} depend on the user characteristics, such as trust bias and conformity bias~\cite{liu2016you,liu2019spiral,agarwal2019addressing,zheng2021disentangling}. 
These specific biases will eventually be coupled into the \textit{data bias} on the user feedback.
In this paper we call this type of data \textit{non-randomized dataset}. 

Since different biases may be coupled, mitigating a set of biases from a data perspective is an important research route.
In addition, it is easier to reduce the system-induced biases by controlling the recommendation policy than by intervening the user to reduce the user-induced biases.
For these reasons, previous works propose to use a special uniform policy to replace the stochastic recommendation policy~\cite{bonner2018causal,liu2020general,chen2021autodebias}. 
Using a uniform policy means that for each user's request, instead of using a recommendation model for item delivery, the system randomly selects some items from all the candidate items, and ranks the selected items with a uniform distribution. 
The users' feedback collected under such a uniform policy is called \textit{randomized dataset}. 
A randomized dataset can be regarded as a good unbiased agent, because it largely avoids the sources of the system-induced biases. 
However, because the uniform logging policy does not take into account each user's preferences and tends to show the users a collection of the items that they are not interested in, it will hurt the users' experiences and the revenue of the platform.
This means that it is necessary to constrain a randomized dataset collection within a particularly limited network traffic.

To utilize such a scarce and precious randomized dataset to help the model training on a non-randomized dataset, the existing methods can be divided into three categories:
1) Use a randomized dataset to re-weight the samples in a non-randomized dataset~\cite{schnabel2016recommendations,yu2020influence}, or to train an imputation model for data augmentation of a non-randomized dataset~\cite{yuan2019improving,liu2020general,zeng2021causal}. In addition, the two can be integrated as a doubly robust framework~\cite{wang2019doubly,chen2021autodebias}.
2) Design a multi-stage training framework to alternately use a non-randomized dataset and a randomized dataset to learn debiased parameters~\cite{wang2021combating,chen2021autodebias}.
3) Use a randomized dataset and a non-randomized log dataset to train two models jointly, and constrain them to be close in some way, so that the model trained on a non-randomized dataset can benefit from the model trained on a randomized dataset~\cite{bonner2018causal,liu2020general}. 
Although these existing works have shown promising results in mitigating the system-induced bias, it is still weak in theoretical insights or an ideal optimization objective function compared with the other more well studied route, i.e., debiased recommendation without a randomized dataset~\cite{wang2020information,saito2020asymmetric,saito2020unbiased,liu2021mitigating}.
This prevents theoretical analysis of the existing methods and a systematic guidance of this research route.

To bridge this gap, we extend previous theoretical insights on debiased recommendation without a randomized dataset~\cite{saito2020asymmetric}.
Specifically, we first formulate a new ideal optimization objective function considering a randomized dataset, and propose a new debiased perspective to facilitate the introduction of some theoretical insights and a more sufficient solution to the system-induced biases, i.e., the debiasing issue is equivalent to directly optimizing the upper bound of this objective function.
Then, we derive two upper bounds of the \textit{unbiased ideal loss function} corresponding to this objective function in practice, i.e., one generalization error bound with the triangle inequality (in Sec.~\ref{subsec:bound_ti}), and the other with the separability (in Sec.~\ref{subsec:bound_s}). 
The difference between the two depends on the different prior constraints satisfied by the adopted loss function. 
We show that most existing methods can be regarded as an insufficient optimization of our upper bound, and propose a novel debiasing method called debiasing approximate upper bound (DUB).
Our method achieves a more sufficient optimization on the upper bound, which is expected to further improve the performance. 
We then conduct extensive experiments on a public dataset and a real product dataset to verify the effectiveness of the proposed method from five different aspects, including unbiased testing scenarios, biased general testing scenarios, the ablation experiments, the distribution of the recommendation lists, and some key factors that may affect the performance of the proposed method. 

The structure of this paper is organised as follows: we briefly introduce some related works in Sec.~\ref{sec:related}; we present some necessary preliminaries in Sec.~\ref{sec:preliminaries}; we give a detailed description of the proposed theoretical insights and method in Sec.~\ref{sec:method}, and discuss the relations to the existing debiasing methods in Sec.~\ref{sec:existing_baselines}; and we analyze and discuss extensive experimental results in Sec.~\ref{sec:empirical}, and present a conclusion and some future directions in Sec.~\ref{sec:feature}.
The contributions of this paper are summarized as follows:
\begin{itemize}
    \item We propose a new debiased perspective and formulate a new ideal optimization objective function with a randomized dataset, based on which a novel solution to the system-induced biases can be obtained by directly minimizing the upper bound of this ideal optimization objective function.
    \item We give some theoretical insights on the upper bound of this ideal optimization objective function, where the adopted loss functions satisfy the triangle inequality and separability, respectively.
    \item We show that most existing solutions can be viewed as an insufficient optimization of the two proposed upper bounds, and then propose a novel method called debiasing approximate upper bound with a randomized dataset (DUB) for a more sufficient optimization of the proposed upper bound.
    \item We conduct extensive experiments on a public dataset and a real product dataset to show the effectiveness of the proposed method, including unbiased evaluation, biased general evaluation, the ablation experiments of the model and the distribution of the recommendation lists, as well as some key factors that may affect the performance of our DUB.
\end{itemize}

\section{Related Work}\label{sec:related}
In this section, we briefly review some related works on two research topics, including debiased recommendation without a randomized dataset and debiased recommendation with a randomized dataset.

\subsection{Debiased Recommendation without A Randomized Dataset} 
Due to the lack of such unbiased guidance information similar to a randomized dataset, the existing works on debiased recommendation without a randomized dataset require making some prior assumptions about the biases, or checking and guaranteeing the unbiasedness of the model based on some specific sophisticated techniques.
The existing works on this research route can be further subdivided into three classes, including heuristic-based methods, inverse propensity score-based methods~\cite{schnabel2016recommendations,yuan2019improving}, and theoretical tools-based methods, depending on the different techniques employed.
A heuristic-based method links a user's feedback with different specific factors to make some prior assumptions about the generation process of some specific biases.
For example, for selection bias in the feedback data (also known as missing not at random mechanism), some previous works have assumed that a user's feedback on an item is related to the user's rating of the item, and a user will only provide his or her own feedback when he or she is particularly satisfied or dissatisfied with the item~\cite{marlin2009collaborative,yang2015boosting}.
In addition to linking with ratings, some subsequent works further consider the different contributions of the user features and the item features in a user's feedback~\cite{kingma2014adam,gopalan2015scalable,liang2016modeling}.
For conformity bias, some previous works assume that a user will use some public opinion as a reference in the process of feedback decision-making, such as hiding or adjusting his or her own feedback~\cite{liu2016you,liu2019spiral,zhang2017modeling,lin2020spiral}.
Based on such prior assumptions, these works usually construct a probabilistic graphical model or a polynomial mixture model containing feature information for a specific bias problem, and then solve the model parameters based on a generalized expectation maximization algorithm.
An inverse propensity score-based method balances the distribution of the items in the observed feedback data by the propensity score estimated based on some variable factors, so that a recommendation model trained on the adjusted non-randomized dataset can avoid the interference of these variable factors as much as possible.
For example, one of the variable factors most often considered in the existing works is the relative exposure frequency of each item in the feedback data, and with the adjustment of the propensity score based on the relative exposure frequency, the exposure distribution of each item in the feedback data is close to uniform~\cite{liang2016modeling,bonner2018causal}.
Moreover, a theoretical tool-based method integrates some theoretical tools from other research fields with debiased recommendation.
They usually derive an unbiased ideal loss function that can be directly optimized for a specific bias problem, or in a case where this unbiased ideal loss function is intractable, further derive a generalization error upper bound for it as a tractable alternative optimization objective.
The common theoretical tools in the existing works include information bottleneck~\cite{wang2020information,liu2021mitigating,liu2022debiased}, positive-unlabeled learning~\cite{saito2020unbiased}, upper bound minimization~\cite{saito2020asymmetric}, disentangled representation learning~\cite{zheng2021disentangling}, and causal inference techniques~\cite{wang2020click,wei2020model}.
Our DUB adopts a similar upper bound minimization idea to provide some new theoretical insights, but is quite different from the previous work~\cite{saito2020asymmetric}.
We propose a new ideal optimization objective function for debiased recommendation with a randomized dataset, whereas the existing works only consider the ideal optimization objective functions defined on a non-randomized dataset.
As described in Sec.~\ref{sec:preliminaries}, this new ideal optimization objective function is more favorable for addressing the system-induced biases.
It can also be seen as an efficient extension of the existing theoretical insights based on upper bound minimization when a randomized dataset is available.
On the other hand, we give more theoretical insights where the prior constraints beyond the triangle inequality are employed to be compatible with more choices of loss functions in practice.

\subsection{Debiased Recommendation with A Randomized Dataset} 
The research on this route additionally introduces a randomized dataset that can act as a proxy for the unbiased information.
Most debiasing methods that fall into this route aim to mine the unbiased knowledge from a randomized dataset by formulating some more sophisticated and efficient techniques, and then use them to guide the training process of a recommendation model on a non-randomized dataset.
The existing works on this research route can be further subdivided into three classes, including inverse propensity score and imputation labels-based methods, multi-stage training-based methods, and joint training-based methods, depending on the different techniques employed.
An inverse propensity score and imputation labels-based method utilizes an additional randomized dataset to estimate the propensity score for each feedback~\cite{schnabel2016recommendations,yu2020influence} or to make the predictions of the imputation labels for unobserved feedback data~\cite{yuan2019improving,liu2020general,liu2022kdcrec,zeng2021causal}.
These obtained propensity scores or imputation labels will be integrated into the model's optimization objective, i.e., transfer the unbiased knowledge into the model's training process.
Propensity score recommendation learning is a representative work in this sub-route, and proposes two methods for the propensity score estimation based on a randomized dataset, including a naive Bayes estimator and a regression model estimator~\cite{schnabel2016recommendations}.
Note that the propensity scores are used in both debiased recommendation routes, and they differ in whether the propensity score is estimated from a non-randomized dataset or a randomized dataset.
In addition, some works also consider estimating and using the propensity scores and imputation labels simultaneously to allow the model to benefit more in a doubly robust framework~\cite{wang2019doubly,chen2021autodebias}.
A multi-stage training-based method designs some effective multi-stage training frameworks in which a non-randomized dataset and a randomized data are used alternately, based on the synergy of which it learns better unbiased parameters.
AutoDebias~\cite{chen2021autodebias} is one of the most representative methods on this research sub-line.
Its main idea is to introduce a meta-learning strategy into a doubly robust debiasing framework to achieve better learning of the model.
Specifically, in each iteration of training, the parameters of the main network (i.e., the recommendation model) in the framework are first fixed, and a randomized dataset is used to better estimate the propensity scores and imputation labels in the auxiliary meta-learning network.
Then, the parameters of the auxiliary meta-learning network are fixed, and a non-randomized dataset is used for unbiased model parameter learning in the main network.
This multi-stage training mode is repeated until the recommendation model converges to a better feasible solution.
Clearly, AutoDebias can be seen as an effective improvement on the training process towards a doubly robust debiasing framework, which is different from most existing debiasing methods that aim to improve the model's optimization objective.
A joint training-based method trains a recommendation model and an auxiliary model for a non-randomized dataset and a randomized dataset, respectively, and uses some custom alignment terms to directly constrain the two models for joint training.
CausE~\cite{bonner2018causal} is a pioneering work of this sub-route and introduces an alignment term of model parameters to facilitate information fusion between the two models.
Since the parameter alignment term will increase the difficulty of model training in a practical application, instead of aligning the two models on the model parameters, Bridge~\cite{liu2020general} constrains the predicted labels of the two models to be as close as possible on an auxiliary set sampled from the full set of feedback.
Different from the existing works, we propose a new perspective on addressing the system-induced biases from the upper bound of an unbiased ideal loss function, and provide a theoretical objective function with a randomized dataset that can be directly optimized.
This means that we convert the task of reducing the system-induced biases to an optimization problem that can be solved directly, which thus provides more guidance on the use of a randomized dataset and the analysis of debiasing methods.

\section{Preliminaries}\label{sec:preliminaries}
\subsection{Notations}\label{subsec:notation}
\begin{figure}[t]
\centering
\includegraphics[width=0.7\columnwidth]{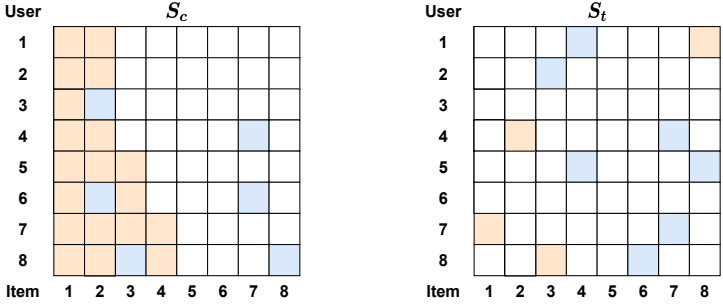}
\caption{An example of the difference between a non-randomized dataset $S_c$ and a randomized dataset $S_t$.}
\label{fig:data}
\end{figure}
A typical recommender system usually takes a user $u_{i}\in U$ as input, and selects an attractive item $v_j\in V$ to be displayed to this user through a stochastic recommendation policy $\pi_c$ deployed by the system, i.e, $v_{j}\sim \pi_{c}\left(\cdot|u_{i}\right)$.
Then, the system will collect the user's feedback on each displayed item $r^{c}_{ij}\sim R^{c}\left(\cdot|u_{i},v_{j}\right)\in \left\{0,1\right\}$, where $r^{c}_{ij}=1$ denotes a positive feedback, $r^{c}_{ij}=0$ denotes a negative feedback, and $R^{c}$ is a complete feedback matrix under $\pi_c$. 
In this paper, we call this type of data \textit{non-randomized dataset} $S_c$.
Based on the collected data $S_c$, the system will retrain a recommendation model $M_c$, and update the recommendation policy. 
Similarly, under a uniform policy $\pi_t$, we have $v_{j}\sim \pi_{t}\left(\cdot|u_{i}\right)$ and $r^{t}_{ij}\sim R^{t}\left(\cdot|u_{i},v_{j}\right)$. 
$R^{t}$ is a complete feedback matrix under $\pi_t$, the feedback of users recorded under $\pi_t$ is called \textit{randomized dataset} $S_t$, and $M_t$ is the auxiliary model trained on $S_t$.

To facilitate understanding of the difference between a non-randomized dataset $S_c$ and a randomized dataset $S_t$, we include an example shown in Figure~\ref{fig:data}, where the recommender system is assumed to contain 8 users and 8 items, and a yellow square and a blue square indicate that the corresponding user-item pair $(u_i,v_j)$ is a positive feedback and a negative feedback, respectively.
Due to the restricted collection process, the scale and scope of $S_t$ are often much smaller than that of $S_c$, where scale refers to the amount of data and scope refers to the coverage of users and items.
We can see from Figure~\ref{fig:data} that in a randomized dataset $S_t$, the number of colored squares is smaller, and there are some users who do not have colored squares.
Due to the nature of a uniform policy $\pi_t$, a randomized dataset $S_t$ suffers from less bias than a non-randomized dataset $S_c$, especially the system-induced biases.
From Figure~\ref{fig:data}, we can see that this relative unbiasedness may be reflected in that each item has a similar probability of getting feedback from different users (i.e., each item has a similar number of colored squares), and each user has a preference distribution that is closer to the ideal state (i.e., due to limited preferences, a user should have far more negative feedback than positive feedback on all items~\cite{pan2008one}).
In addition, we can also see from Figure~\ref{fig:data} that a randomized dataset $S_t$ may reveal interests for a user that are not perceived in a recommendation policy $\pi_c$, such as user 1 for item 8, and may correct for pseudo-negative feedback in a non-randomized dataset $S_c$ subject to the system-induced biases, such as user 8 for item 3.
Note that in order to ensure non-overlapping between $S_c$ and $S_t$, and because the feedback data in $S_t$ is more unbiased and credible, we actually remove from $S_c$ those feedback data that appear in $S_t$,  such as user 8 for item 3.

Since a non-randomized dataset $S_c$ and a randomized dataset $S_t$ are part of the complete feedback matrix (i.e., $R^c$ and $R^t$) under a recommendation policy $\pi_c$ and a uniform policy $\pi_t$, respectively, we can intuitively think that $R^c$ and $R^t$ inherit this difference in bias between $S_c$ and $S_t$, i.e., $R^t$ has a better unbiasedness than $R^c$.
In particular, each element in $R^t$ can be thought of as a user's feedback result after an item has been displayed in all possible ways.
This is a result that can be gradually achieved through the long-term deployment of a uniform policy $\pi_t$.
Unlike $R^t$, even if we can obtain $R^c$, it can only alleviate some of the biases induced by the system, and still inevitably suffers from the rest of these biases, especially pseudo-negative feedback.

\subsection{Problem Formulation}
The optimization objective of most existing recommendation methods is the average loss function over the observed feedback under a policy $\pi_c$,
\begin{equation}\label{equ:naive_estimator}
\mathcal{L}_{observed}^{\ell}\left(R^c, \hat{R}^c\right) = \frac{1}{\left|\mathcal{O}\right|}\sum_{\left(i,j\right)\in \mathcal{O}}\ell \left(R_{i,j}^c,\hat{R}_{i,j}^c\right),
\end{equation}
where $\mathcal{O}\in\left\{\left(i,j\right)\right\}$ denotes a set of observed feedback. $\hat{R}^c$ denotes the predicted label matrix of $M_c$, and $\ell\left(\cdot, \cdot\right)$ is an arbitrary loss function. 
Eq.(\ref{equ:naive_estimator}) can be regarded as the simplest estimator of the ideal optimization objective under policy $\pi_c$, 
\begin{equation}\label{equ:sc_ideal_estimator}
\mathcal{L}_{{\pi_c}-{ideal}}^{\ell}\left(R^c, \hat{R}^c\right) = \frac{1}{\left|\mathcal{D}\right|}\sum_{\left(i,j\right)\in \mathcal{D}}\ell \left(R_{i,j}^c,\hat{R}_{i,j}^c\right),
\end{equation}
where $\mathcal{D}$ denotes the complete set of feedback. 
Due to the system-induced biases, Eq.(\ref{equ:naive_estimator}) is not an unbiased estimation of Eq.(\ref{equ:sc_ideal_estimator})~\cite{marlin2009collaborative,steck2013evaluation}. 
Instead, some previous works on debiased recommendation without a randomized dataset have shown that better performance can be obtained by optimizing an unbiased estimation or a generalization error bound of Eq.(\ref{equ:sc_ideal_estimator})~\cite{schnabel2016recommendations,saito2020asymmetric}.

However, as described in Sec.~\ref{subsec:notation}, even if we can obtain the complete feedback matrix $R^c$ under a recommendation policy $\pi_c$, $R^c$ can only alleviate some but not all of the biases induced by the system.
This means that an unbiased estimator for $R^c$ is not necessarily equivalent to an ideal unbiased evaluation.
To further solve the system-induced biases, based on the analysis in Sec~\ref{subsec:notation}, we argue that a better option is to use $R^t$ instead of $R^c$.
This is because $R^t$ consisting of a randomized data $S_t$ obviously contains better relative unbiasedness than $R^c$. 
Based on this idea, we formulate a new \textit{ideal optimization objective function},
\begin{equation}\label{equ:st_ideal_estimator}
\mathcal{L}_{{\pi_t}-{ideal}}^{\ell}\left(R^t, \hat{R}^c\right) = \frac{1}{\left|\mathcal{D}\right|}\sum_{\left(i,j\right)\in \mathcal{D}}\ell \left(R_{i,j}^t,\hat{R}_{i,j}^c\right).
\end{equation}
This means that we can optimize Eq.(\ref{equ:st_ideal_estimator}) as a better solution to the system-induced bias problem, and Eq.(\ref{equ:st_ideal_estimator}) can also be seen as an efficient extension of the existing ideal optimization objective functions when a randomized dataset is available.
However, it is very difficult to directly optimize Eq.(\ref{equ:st_ideal_estimator}).
On one hand, we only have a small part of the real feedback of $R^t$, i.e., $S_t$. 
On the other hand, although we have a non-randomized dataset $S_c$, we do not know the corresponding feedback in $R^t$ for these feedback data, i.e., the state of a non-randomized dataset $S_c$ in $R^t$ is unknown.
In particular, we need to answer the following question: If the items in $S_c$ are randomly displayed, what will the feedback be like? 
This involves the concept of counterfactual, which is recognized as a challenging problem~\cite{pearl2018book}. 
To address this challenge, we will turn to deriving an upper bound of Eq.(\ref{equ:st_ideal_estimator}), and propose a general debiasing framework based on upper bound minimization, where the upper bound of Eq.(\ref{equ:st_ideal_estimator}) will be taken as a new optimization objective function to drive a tractable solution.

\section{The Proposed Method}\label{sec:method}
In this section, we first present some theoretical insights into debiased recommendation with a randomized dataset.
Specifically, our goal is to derive an upper bound of the \textit{ideal optimization objective function} in Eq.(\ref{equ:st_ideal_estimator}) by extending the theory in~\cite{saito2020asymmetric}, and use it as an alternative objective that can be directly optimized.
Note that in practice, we need to specify the type of loss function $\ell$ in this optimization objective, and we refer to the objective function having a specific form as the \textit{unbiased ideal loss function} in the following.
Different types of loss functions satisfy different prior constraints and have different effects on theoretical insights.
Therefore, in order to be compatible with as many types of loss functions as possible, we propose two corresponding upper bounds when the adopted loss functions $\ell$ satisfy the triangular inequality (in Sec.~\ref{subsec:bound_ti}) and separability (in Sec.~\ref{subsec:bound_s}), respectively.
Secondly, we discuss the generalization error bounds to clarify the key factors. 
Finally, we give a detailed description of the proposed method, i.e., DUB. 
Note that unless otherwise specified, we abbreviate $\mathcal{L}_{{\pi_t}-ideal}^{\ell}\left(R^t, \hat{R}^c\right)$ as $\mathcal{L}\left(R^t, \hat{R}^c\right)$ in the following for brevity. 
For ease of reference, the main notations in theoretical analysis are listed in Table~\ref{table:natations}.
\begin{table}[htbp]
	\centering
	\caption{The main notations and explanations.}
	\label{table:natations}
	\scalebox{1.0}{
	\begin{tabular}{l|l}
	\specialrule{0.1em}{3pt}{3pt}
	    \textbf{Symbol} & \textbf{Meaning}\\
		\specialrule{0.05em}{3pt}{3pt}

            $S_t$ & a randomized dataset\\
            
            $S_c$ & a non-randomized dataset\\ 
            
            $S_u$ & the unobserved data\\
            
		    $\mathcal{D}$ & the whole set of data, i.e., $\mathcal{D} = S_c \cup S_t \cup S_u$\\
            
            $M_c$ & the recommendation model trained on a non-randomized dataset $S_c$\\
            
            $M_t$ & the auxiliary model trained on a randomized dataset $S_t$\\
            
            $R^{*}$ & the complete feedback matrix under $\pi_{*}$, $\pi_{*} \in \left\{\pi_c,\pi_t\right\}$\\
            
            $\hat{R}^{*}$ & the predicted label matrix of $M_{*}$, $M_{*} \in \left\{M_c,M_t\right\}$\\
            
            \specialrule{0.05em}{3pt}{3pt}
            
            \multirow{2}{*}{$I_{S_{*}}$}  & the set of user-item pair indices contained in the feedback data $S_{*}$,
            \\&
            where $S_{*} \in \left\{S_c,S_t,S_u\right\}$\\
            
            \specialrule{0.05em}{3pt}{3pt}
            
            $\mathcal{L}\left(R^t,\hat{R}^c\right)$ & the unbiased ideal loss function when a randomized dataset is available\\
            
            \specialrule{0.05em}{3pt}{3pt}
            
            \multirow{3}{*}{$\mathcal{L}^{S_{*}}\left(\cdot, \cdot\right)$} & the loss function defined on the set of user-item pair indices contained in
            \\& 
            the feedback data $S_{*}$ with the size of the whole set as the denominator, i.e.,
            \\& 
            $\mathcal{L}^{S_{*}}\left(\cdot, \cdot\right)=\frac{1}{\left|\mathcal{D}\right|}\sum_{\left(u,v\right)\in I_{S_{*}}}\ell \left(\cdot, \cdot\right)$\\
            
            \specialrule{0.05em}{3pt}{3pt}
            
            \multirow{2}{*}{$\mathcal{L}_{|S_{*}|}^{S_{*}}\left(\cdot, \cdot\right)$} & the average loss function defined on the set of user-item pair indices contained
            \\& 
            in the feedback data $S_{*}$, i.e., $\mathcal{L}_{|S_{*}|}^{S_{*}}\left(\cdot, \cdot\right)=\frac{1}{\left|S_{*}\right|}\sum_{\left(u,v\right)\in I_{S_{*}}}\ell \left(\cdot, \cdot\right)$\\
            
	\specialrule{0.1em}{3pt}{3pt}
	\end{tabular}}
\end{table}

In order to emphasize a confusing notation $\mathcal{L}^{S_{*}}\left(\cdot, \cdot\right)$, we further describe the difference between $\mathcal{L}^{S_c}\left(R^c,\hat{R}^c\right)$, $\mathcal{L}^{S_c}\left(R^c,\hat{R}^t\right)$ and $\mathcal{L}^{S_c}\left(R^t,\hat{R}^t\right)$ as an example.
By definition, $\mathcal{L}^{S_c}\left(R^c,\hat{R}^c\right)$ denotes a loss function defined on the set of user-item pair indices contained in the feedback data $S_c$.
Therefore, the true labels used in this loss function are the corresponding part of $R^c$ on the specific user-item pair index set $I_{S_c}$.
Obviously, the true labels at this time are the feedback labels of a non-randomized dataset $S_c$.
Similarly, the predicted labels used in the loss function are the predicted outputs of the recommendation model $M_c$ for each sample in a non-randomized dataset $S_c$.
For $\mathcal{L}^{S_c}\left(R^c,\hat{R}^t\right)$, the true labels used are also the feedback labels of a non-randomized dataset $S_c$, but the predicted labels used are changed to the part of $\hat{R}^t$ on the specific user-item pair index set $I_{S_c}$, i.e., the predicted outputs of the auxiliary model $M_t$ for each sample in a non-randomized dataset $S_c$.
In particular, for $\mathcal{L}^{S_c}\left(R^t,\hat{R}^t\right)$, the true labels used in the loss function are changed to the part of $R^t$ on the specific user-item pair index set $I_{S_c}$.
Obviously, as described in Sec.~\ref{subsec:notation}, we cannot know the true labels of this part of the feedback data in practice, i.e., it cannot be optimized directly using the supervision information.

\subsection{Theoretical Analysis}
\subsubsection{A Generalization Error Bound with the Triangle Inequality}\label{subsec:bound_ti}
Similar to most works using the upper bound minimization framework~\cite{courty2017joint,saito2020asymmetric}, we first consider the case when the adopted loss function $\ell$ satisfies the triangle inequality, e.g., the 0-1 loss and $l_1$-norm~\cite{lee2019domain,kaspar2019neural}.
In Proposition~\ref{prop:triangle}, we first derive a simple upper bound on Eq.(\ref{equ:st_ideal_estimator}) based on this prior constraint.
\begin{myprop}{4.1}\label{prop:triangle}
Assume that the loss function $\ell$ obeys the triangle inequality. Then, for any given predicted label matrices $\hat{R}^t$ and $\hat{R}^c$, the following inequality holds.
    \begin{align}
        \mathcal{L}\left(R^t,\hat{R}^c\right)&\leq \mathcal{L}^{S_t}\left(R^t,\hat{R}^c\right) + 
        \mathcal{L}^{S_c}\left(R^t,R^c\right)
        +\mathcal{L}^{S_c}\left(R^c,\hat{R}^c\right)+\mathcal{L}^{S_u}\left(R^t,\hat{R}^t\right) + \mathcal{L}^{S_u}\left(\hat{R}^t,\hat{R}^c\right).\notag
    \end{align}
\end{myprop}
\begin{proof}
    \begin{align}
        \mathcal{L}\left(R^t,\hat{R}^c\right) &=\mathcal{L}^{S_t}\left(R^t,\hat{R}^c\right) + \mathcal{L}^{S_c}\left(R^t,\hat{R}^c\right) + 
        \mathcal{L}^{S_u}\left(R^t,\hat{R}^c\right)\notag\\
        &\leq \mathcal{L}^{S_t}\left(R^t,\hat{R}^c\right)  
        +\mathcal{L}^{S_c}\left(R^t,R^c\right)+
        \mathcal{L}^{S_c}\left(R^c,\hat{R}^c\right)+\mathcal{L}^{S_u}\left(R^t,\hat{R}^t\right) + \mathcal{L}^{S_u}\left(\hat{R}^t,\hat{R}^c\right).\notag
    \end{align}
where $S_u$ denotes the set of unobserved feedback, i.e., $\mathcal{D}=S_c\cup S_t\cup S_u$, $\mathcal{L}^{S_{*}}\left(\cdot, \cdot\right)=\frac{1}{\left|\mathcal{D}\right|}\sum_{\left(u,v\right)\in I_{S_{*}}}\ell \left(\cdot, \cdot\right)$, and $S_{*}\in\left\{S_c,S_t,S_u\right\}$. 
We first divide Eq.(\ref{equ:st_ideal_estimator}) into a summation of three disjoint subsets, and apply the triangle inequality to $\mathcal{L}^{S_c}\left(R^t,\hat{R}^c\right)$ and $\mathcal{L}^{S_u}\left(R^t,\hat{R}^c\right)$. 
Note that as described in Sec.~\ref{subsec:notation}, the disjoint properties of $S_c$ and $S_t$ are ensured during the data collection phase.
\end{proof}

The fourth term in Proposition~\ref{prop:triangle} is difficult to be solved because we only know the true labels of a small part of $R_t$, i.e., $S_t$, but not the true labels of $R_t$ on the specific user-item pair index set $I_{S_u}$.
Therefore, through the Hoeffding's inequality~\cite{hoeffding1994probability}, we convert it into an easy-to-solve alternative, and further analyze the generalization error bound of the unbiased ideal loss function.

\begin{mytheo}{4.2}[Generalization Error Bound of Unbiased Ideal Loss \uppercase\expandafter{\romannumeral1}]\label{theorem:triangle}
Assume that two predicted matrices $\hat{R}^t$ and $\hat{R}^c$ are given, and a loss function $\ell$ obeys the triangle inequality and is bounded by a positive constant $\Delta$. Then, for any finite hypothesis space of predictions $\mathcal{H}=\left\{\hat{R}_1^t,\cdots,\hat{R}_{\left|\mathcal{H}\right|}^t\right\}$, and for any $\eta\in\left(0,1\right)$, the ideal loss $\mathcal{L}\left(R^t,\hat{R}^c\right)$ is bounded with probability $1-\eta$ by:
     \begin{align}\label{equ:triangle}
        \mathcal{L}\left(R^t,\hat{R}^c\right)\notag
        &\leq \underbrace{\mathcal{L}^{S_t}\left(R^t,\hat{R}^c\right)}_{(a)}
        +\underbrace{\mathcal{L}^{S_c}\left(R^t,R^c\right)}_{(b)}
        +\underbrace{\mathcal{L}^{S_c}\left(R^c,\hat{R}^c\right)}_{(c)}\notag\\
        &+\underbrace{\mathcal{L}^{S_u}\left(\hat{R}^t,\hat{R}^c\right)}_{(d)}+\underbrace{\mathcal{L}_{|S_t|}^{S_t}\left(R^t,\hat{R}^t\right)}_{(e)}+bias\left(\mathcal{L}_{|S_t|}^{S_t}\left(R^t,\hat{R}^t\right)\right)\notag\\
        &+\frac{\Delta}{\left|S_t\right|}\sqrt{\frac{\left|\mathcal{D}\right|}{2}\log\left(\frac{2\left|\mathcal{H}\right|}{\eta}\right)},
    \end{align}
where  $bias\left(\mathcal{L}_{|S_t|}^{S_t}\left(R^t,\hat{R}^t\right)\right)=\mathcal{L}^{S_u}\left(R^t,\hat{R}^t\right)-\mathbb{E}\left[\mathcal{L}_{|S_t|}^{S_t}\left(R^t,\hat{R}^t\right)\right]$ is the error term caused by using $\mathbb{E}\left[\mathcal{L}_{|S_t|}^{S_t}\left(R^t,\hat{R}^t\right)\right]$ to replace $\mathcal{L}^{S_u}\left(R^t,\hat{R}^t\right)$, and ${\mathcal{L}_{|S_t|}^{S_t}\left(R^t,\hat{R}^t\right)}=\frac{1}{\left|S_t\right|}\sum_{\left(i,j\right)\in S_t}\ell \left(R_{i,j}^t,\hat{R}_{i,j}^t\right)$.
\end{mytheo}

\begin{proof}
Our goal is to use the easy-to-solve term $(e)$ in Eq.(\ref{equ:triangle}) to replace the fourth difficult-to-solve term in Proposition~\ref{prop:triangle}, and obtain the approximate error term corresponding to this operation.

First, we have the following equation:
    \begin{align}
    \label{eq:2}
        \mathcal{L}^{S_u}\left(R^t,\hat{R}^t\right)&\notag=\mathcal{L}^{S_u}\left(R^t,\hat{R}^t\right)-\mathbb{E}\left[\mathcal{L}_{|S_t|}^{S_t}\left(R^t,\hat{R}^t\right)\right]+\mathbb{E}\left[\mathcal{L}_{|S_t|}^{S_t}\left(R^t,\hat{R}^t\right)\right]\notag\\
        &=\mathbb{E}\left[\mathcal{L}_{|S_t|}^{S_t}\left(R^t,\hat{R}^t\right)\right]+bias\left(\mathcal{L}_{|S_t|}^{S_t}\left(R^t,\hat{R}^t\right)\right).
    \end{align}
Using the Hoeffding's inequality and union bounds to make a uniform convergence argument, we get: 
    \begin{align}
        &P\left(\left|\mathbb{E}\left[\mathcal{L}_{|S_t|}^{S_t}\left(R^t,\hat{R}^t\right)\right]-\mathcal{L}_{|S_t|}^{S_t}\left(R^t,\hat{R}^t\right)\right|\leq\epsilon\right)\geq 1-\eta\notag\\
        &\Leftarrow P\left(\max_{\hat{R}_h^t\in\mathcal{H}}\left|\mathbb{E}\left[\mathcal{L}_{|S_t|}^{S_t}\left(R^t,\hat{R}_h^t\right)\right]-\mathcal{L}_{|S_t|}^{S_t}\left(R^t,\hat{R}_h^t\right)\right|\leq\epsilon\right)\geq 1-\eta\notag\\
        &\Leftrightarrow P\left(\bigcup_{\hat{R}_h^t\in\mathcal{H}}^{}\left|\mathbb{E}\left[\mathcal{L}_{|S_t|}^{S_t}\left(R^t,\hat{R}_h^t\right)\right]-\mathcal{L}_{|S_t|}^{S_t}\left(R^t,\hat{R}_h^t\right)\right|\geq\epsilon\right)\leq\eta\notag\\
        &\Leftarrow\sum_{h=1}^{\left|\mathcal{H}\right|}P\left(\left|\mathbb{E}\left[\mathcal{L}_{|S_t|}^{S_t}\left(R^t,\hat{R}_h^t\right)\right]-\mathcal{L}_{|S_t|}^{S_t}\left(R^t,\hat{R}_h^t\right)\right|\geq\epsilon\right)\leq\eta\notag\\
        &\Leftarrow\left|\mathcal{H}\right|\times2\exp\left(\frac{-2\left|S_t\right|^{2}\epsilon^2}{\left|\mathcal{D}\right|\Delta^2}\right)\leq\eta\notag.
    \end{align}
Solving for $\epsilon$ yields the bound
    \begin{align}
    \label{eq:3}
        &\left|\mathbb{E}\left[\mathcal{L}_{|S_t|}^{S_t}\left(R^t,\hat{R}^t\right)\right]-\mathcal{L}_{|S_t|}^{S_t}\left(R^t,\hat{R}^t\right)\right|\leq\frac{\Delta}{\left|S_t\right|}\sqrt{\frac{\left|\mathcal{D}\right|}{2}\log\left(\frac{2\left|\mathcal{H}\right|}{\eta}\right)}\notag\\
        &\Rightarrow\mathbb{E}\left[\mathcal{L}_{|S_t|}^{S_t}\left(R^t,\hat{R}^t\right)\right]\leq\mathcal{L}_{|S_t|}^{S_t}\left(R^t,\hat{R}^t\right)+\frac{\Delta}{\left|S_t\right|}\sqrt{\frac{\left|\mathcal{D}\right|}{2}\log\left(\frac{2\left|\mathcal{H}\right|}{\eta}\right)}.
    \end{align}
By combining Eq.(\ref{eq:2}) and Eq.(\ref{eq:3}), we get the following inequality, which holds with a probability of at least $1-\eta$:
    \begin{align}
    \label{eq:1}
        \mathcal{L}^{S_u}\left(R^t,\hat{R}^t\right)&\leq\mathcal{L}_{|S_t|}^{S_t}\left(R^t,\hat{R}^t\right)+\frac{\Delta}{\left|S_t\right|}\sqrt{\frac{\left|\mathcal{D}\right|}{2}\log\left(\frac{2\left|\mathcal{H}\right|}{\eta}\right)}+bias\left(\mathcal{L}_{|S_t|}^{S_t}\left(R^t,\hat{R}^t\right)\right).
    \end{align}
Then, by combining Proposition~\ref{prop:triangle} and Eq.(\ref{eq:1}), the proof is completed.
\end{proof}

\subsubsection{A Generalization Error Bound with the Separability}\label{subsec:bound_s}
Note that in recommender systems, some widely used loss functions do not satisfy the triangular inequality, e.g., the cross-entropy loss and the mean square error. To further expand the optional range of the loss function, we propose a new prior constraint on the loss function,
\begin{mydef}{4.3}
\textbf{Separability}: A loss is considered to satisfy the separability if and only if the following inequality holds,
    \begin{align}
        \mathcal{L}^{\ell}\left(c,a\right)\leq\mathcal{L}^{\ell}\left(b,a\right)+\mathcal{L}^{\ell}\left(c-b,a\right).\notag
    \end{align}
\end{mydef}
\begin{proof}
As an example, we prove that the binary cross-entropy loss satisfies the separability, and other loss functions can be checked in a similar process.
Given a form of the binary cross-entropy loss $\mathcal{L}^{\ell} \left(y,\hat{y}\right) =-\left[y\log\hat{y}+(1-y)\log(1-\hat{y})\right]$, where $y\in\left\{0,1\right\}$, we can derive that 
\begin{align}
    \mathcal{L}^{\ell} (c,a)-\mathcal{L}^{\ell} (b,a)\notag
    &= -\left[ c\log a+(1-c)\log(1-a)\right] + [ b\log a\\\notag
    &\quad+(1-b)\log(1-a)] \\\notag
    & = -\left[(c-b)\log a  - (c-b)\log (1-a)\right]\\\notag
    &\leq -\left[(c-b)\log a  - (c-b)\log (1-a)\right]\\\notag
    &\quad-\log(1-a)\\\notag
    & = -\left[(c-b)\log a  + (1-(c-b))\log(1-a)\right] \\\notag
    &= \mathcal{L}^{\ell} (c-b,a ).\notag
\end{align}
The inequality conversion in the process can be obtained because of the non-negativity of $-\log(1-a)$, where $0\leq a\leq1$. Then, the binary cross-entropy loss satisfies the separability. 
\end{proof}

Based on the separability, similar to the proof process of Proposition~\ref{prop:triangle} and Theory~\ref{theorem:triangle}, we can get Proposition~\ref{prop:separability} and Theory~\ref{theorem:separability}.

\begin{myprop}{4.4}\label{prop:separability}
Assume that the loss function $\ell$ obeys the separability. Then, for any given predicted label matrices $\hat{R}_t$ and $\hat{R}_c$, the following inequality holds.
    \begin{align}
        \mathcal{L}\left(R^t,\hat{R}^c\right)&\leq \mathcal{L}^{S_t}\left(R^t,\hat{R}^c\right) + 
        \mathcal{L}^{S_c}\left(R^t-R^c,\hat{R}^c\right)\notag+\mathcal{L}^{S_c}\left(R^c,\hat{R}^c\right) \\
        &+\mathcal{L}^{S_u}\left(R^t-\hat{R}^t,\hat{R}^c\right) + \mathcal{L}^{S_u}\left(\hat{R}^t,\hat{R}^c\right).\notag
    \end{align}
\end{myprop}
\begin{proof}
    \begin{align}
        \mathcal{L}\left(R^t,\hat{R}^c\right) &=\mathcal{L}^{S_t}\left(R^t,\hat{R}^c\right) + \mathcal{L}^{S_c}\left(R^t,\hat{R}^c\right) + 
        \mathcal{L}^{S_u}\left(R^t,\hat{R}^c\right)\notag\\
        &\leq \mathcal{L}^{S_t}\left(R^t,\hat{R}^c\right) + 
        \mathcal{L}^{S_c}\left(R^t-R^c,\hat{R}^c\right)\notag+\mathcal{L}^{S_c}\left(R^c,\hat{R}^c\right) \\
        &+\mathcal{L}^{S_u}\left(R^t-\hat{R}^t,\hat{R}^c\right) + \mathcal{L}^{S_u}\left(\hat{R}^t,\hat{R}^c\right).\notag
    \end{align}
where we apply the separability to $\mathcal{L}^{S_c}\left(R^t,\hat{R}^c\right)$ and $\mathcal{L}^{S_u}\left(R^t,\hat{R}^c\right)$.
\end{proof}

\begin{mytheo}{4.5}[Generalization Error Bound of Unbiased Ideal Loss \uppercase\expandafter{\romannumeral2}]\label{theorem:separability}
Assume that two predicted matrices $\hat{R}^t$ and $\hat{R}^c$ are given, and a loss function $\ell$ obeys the separability and is bounded by a positive constant $\Delta$. Then, for any finite hypothesis space of predictions $\mathcal{H}=\left\{\hat{R}_1^c,\cdots,\hat{R}_{\left|\mathcal{H}\right|}^c\right\}$, and for any $\eta\in\left(0,1\right)$, the ideal loss $\mathcal{L}\left(R^t,\hat{R}^c\right)$ is bounded with probability $1-\eta$ by:
    \begin{align}\label{equ:separability}
        \mathcal{L}\left(R^t,\hat{R}^c\right)
        &\leq \underbrace{\mathcal{L}^{S_t}\left(R^t,\hat{R}^c\right)}_{(a)}
        +\underbrace{\mathcal{L}^{S_c}\left(R^t-R^c,\hat{R}^c\right)}_{(b)}
        +\underbrace{\mathcal{L}^{S_c}\left(R^c,\hat{R}^c\right)}_{(c)}\notag\\
        &+\underbrace{\mathcal{L}^{S_u}\left(\hat{R}^t,\hat{R}^c\right)}_{(d)}+\underbrace{\mathcal{L}_{|S_t|}^{S_t}\left(R^t-\hat{R}^t,\hat{R}^c\right)}_{(e)}\notag\\
        &+\frac{\Delta}{\left|S_t\right|}\sqrt{\frac{\left|\mathcal{D}\right|}{2}\log\left(\frac{2\left|\mathcal{H}\right|}{\eta}\right)} +bias\left(\mathcal{L}_{|S_t|}^{S_t}\left(R^t-\hat{R}^t,\hat{R}^c\right)\right).
    \end{align}
\end{mytheo}
\begin{proof}
Our goal is to use the easy-to-solve term $(e)$ in Eq.(\ref{equ:separability}) to replace the fourth difficult-to-solve term in Proposition~\ref{prop:separability}, and obtain the approximate error term corresponding to this operation.

First, we have the following equation:
    \begin{align}
    \label{eq:2_}
        \mathcal{L}^{S_u}\left(R^t-\hat{R}^t,\hat{R}^c\right)&\notag=\mathcal{L}^{S_u}\left(R^t-\hat{R}^t,\hat{R}^c\right)-\mathbb{E}\left[\mathcal{L}_{|S_t|}^{S_t}\left(R^t-\hat{R}^t,\hat{R}^c\right)\right]+\mathbb{E}\left[\mathcal{L}_{|S_t|}^{S_t}\left(R^t-\hat{R}^t,\hat{R}^c\right)\right]\notag\\
        &=\mathbb{E}\left[\mathcal{L}_{|S_t|}^{S_t}\left(R^t-\hat{R}^t,\hat{R}^c\right)\right]+bias\left(\mathcal{L}_{|S_t|}^{S_t}\left(R^t-\hat{R}^t,\hat{R}^c\right)\right).
    \end{align}
Using the Hoeffding's inequality and union bounds to make a uniform convergence argument, we get: 
    \begin{align}
        &P\left(\left|\mathbb{E}\left[\mathcal{L}_{|S_t|}^{S_t}\left(R^t-\hat{R}^t,\hat{R}^c\right)\right]-\mathcal{L}_{|S_t|}^{S_t}\left(R^t-\hat{R}^t,\hat{R}^c\right)\right|\leq\epsilon\right)\geq 1-\eta\notag\\
        &\Leftarrow P\left(\max_{\hat{R}_h^c\in\mathcal{H}}\left|\mathbb{E}\left[\mathcal{L}_{|S_t|}^{S_t}\left(R^t-\hat{R}^t,\hat{R}_h^c\right)\right]-\mathcal{L}_{|S_t|}^{S_t}\left(R^t-\hat{R}^t,\hat{R}_h^c\right)\right|\leq\epsilon\right)\geq 1-\eta\notag\\
        &\Leftrightarrow P\left(\bigcup_{\hat{R}_h^c\in\mathcal{H}}^{}\left|\mathbb{E}\left[\mathcal{L}_{|S_t|}^{S_t}\left(R^t-\hat{R}^t,\hat{R}_h^c\right)\right]-\mathcal{L}_{|S_t|}^{S_t}\left(R^t-\hat{R}^t,\hat{R}_h^c\right)\right|\geq\epsilon\right)\leq\eta\notag\\
        &\Leftarrow\sum_{h=1}^{\left|\mathcal{H}\right|}P\left(\left|\mathbb{E}\left[\mathcal{L}_{|S_t|}^{S_t}\left(R^t-\hat{R}^t,\hat{R}_h^c\right)\right]-\mathcal{L}_{|S_t|}^{S_t}\left(R^t-\hat{R}^t,\hat{R}_h^c\right)\right|\geq\epsilon\right)\leq\eta\notag\\
        &\Leftarrow\left|\mathcal{H}\right|\times2\exp\left(\frac{-2\left|S_t\right|^{2}\epsilon^2}{\left|\mathcal{D}\right|\Delta^2}\right)\leq\eta\notag.
    \end{align}
Solving for $\epsilon$ yields the bound
    \begin{align}
    \label{eq:3_}
        &\left|\mathbb{E}\left[\mathcal{L}_{|S_t|}^{S_t}\left(R^t-\hat{R}^t,\hat{R}^c\right)\right]-\mathcal{L}_{|S_t|}^{S_t}\left(R^t-\hat{R}^t,\hat{R}^c\right)\right|\leq\frac{\Delta}{\left|S_t\right|}\sqrt{\frac{\left|\mathcal{D}\right|}{2}\log\left(\frac{2\left|\mathcal{H}\right|}{\eta}\right)}\notag\\
        &\Rightarrow\mathbb{E}\left[\mathcal{L}_{|S_t|}^{S_t}\left(R^t-\hat{R}^t,\hat{R}^c\right)\right]\leq\mathcal{L}_{|S_t|}^{S_t}\left(R^t-\hat{R}^t,\hat{R}^c\right)+\frac{\Delta}{\left|S_t\right|}\sqrt{\frac{\left|\mathcal{D}\right|}{2}\log\left(\frac{2\left|\mathcal{H}\right|}{\eta}\right)}.
    \end{align}
By combining Eq.(\ref{eq:2_}) and Eq.(\ref{eq:3_}), we get the following inequality, which holds with a probability of at least $1-\eta$:
    \begin{align}
    \label{eq:1_}
        \mathcal{L}^{S_u}\left(R^t-\hat{R}^t,\hat{R}^c\right)&\leq\mathcal{L}_{|S_t|}^{S_t}\left(R^t-\hat{R}^t,\hat{R}^c\right)+\frac{\Delta}{\left|S_t\right|}\sqrt{\frac{\left|\mathcal{D}\right|}{2}\log\left(\frac{2\left|\mathcal{H}\right|}{\eta}\right)}+bias\left(\mathcal{L}_{|S_t|}^{S_t}\left(R^t-\hat{R}^t,\hat{R}^c\right)\right).
    \end{align}
Then, by combining Proposition~\ref{prop:separability} and Eq.(\ref{eq:1_}), the proof is completed.
\end{proof}

\subsection{Analysis of the Generalization Error Bounds}\label{subsec:analysis}
As suggested in Theory~\ref{theorem:triangle} and Theory~\ref{theorem:separability}, we list the corresponding explanation for each term in the generalization error bounds. 
For different terms in the two generalization error bounds, we use indexes 1 and 2 to denote the upper bound of the triangle inequality and the upper bound of the separability, respectively.
The two generalization error bounds are the same in terms $(a)$, $(c)$, and $(d)$, but are different in terms $(b)$ and $(e)$. 
\begin{enumerate}
    \item[(a)] By definition, $\mathcal{L}^{S_t}\left(R^t,\hat{R}^c\right)=\frac{1}{\left|\mathcal{D}\right|}\sum_{\left(u,v\right)\in I_{S_t}}\ell \left(R^t,\hat{R}^c\right)$, i.e., the predicted loss of $M_c$ with the size of the whole set as the denominator w.r.t. the true feedback labels on $S_t$.
    \item[(b.1)] By definition, $\mathcal{L}^{S_c}\left(R^t,R^c\right)=\frac{1}{\left|\mathcal{D}\right|}\sum_{\left(u,v\right)\in I_{S_c}}\ell \left(R^t,R^c\right)$, i.e., the difference between the true feedback labels of policy $\pi_c$ and policy $\pi_t$ on the specific user-item pair index set $I_{S_c}$.
    \item[(b.2)] By definition, $\mathcal{L}^{S_c}\left(R^t-R^c,\hat{R}^c\right)=\frac{1}{\left|\mathcal{D}\right|}\sum_{\left(u,v\right)\in I_{S_c}}\ell \left(R^t-R^c,\hat{R}^c\right)$, i.e., the predicted loss of $M_c$ w.r.t. the difference between the true feedback labels of policy $\pi_c$ and policy $\pi_t$ on the specific user-item pair index set $I_{S_c}$.
    \item[(c)] By definition, $\mathcal{L}^{S_c}\left(R^c,\hat{R}^c\right)=\frac{1}{\left|\mathcal{D}\right|}\sum_{\left(u,v\right)\in I_{S_c}}\ell \left(R^c,\hat{R}^c\right)$, i.e., the supervised loss of $M_c$ with the size of the whole set as the denominator w.r.t. the true feedback labels on $S_c$.
    \item[(d)] By definition, $\mathcal{L}^{S_u}\left(\hat{R}^t,\hat{R}^c\right)=\frac{1}{\left|\mathcal{D}\right|}\sum_{\left(u,v\right)\in I_{S_u}}\ell \left(\hat{R}^t,\hat{R}^c\right)$, i.e., the unsupervised loss between $M_t$ and $M_c$ on the specific user-item pair index set $I_{S_u}$.
    \item[(e.1)] By definition, $\mathcal{L}_{|S_t|}^{S_t}\left(R^t,\hat{R}^t\right)=\frac{1}{\left|S_t\right|}\sum_{\left(u,v\right)\in I_{S_t}}\ell \left(R^t,\hat{R}^t\right)$, i.e., the supervised loss of $M_t$ w.r.t. the true feedback labels on $S_t$.
    \item[(e.2)] By definition, $\mathcal{L}_{|S_t|}^{S_t}\left(R^t-\hat{R}^t,\hat{R}^c\right)=\frac{1}{\left|S_t\right|}\sum_{\left(u,v\right)\in I_{S_t}}\ell \left(R^t-\hat{R}^t,\hat{R}^c\right)$, i.e., the predicted loss of $M_c$ w.r.t. the prediction error of $M_t$ on the specific user-item pair index set $I_{S_t}$.
\end{enumerate}

Intuitively, the three common terms $(a)$, $(c)$ and $(d)$ can be viewed as the supervised loss of $M_c$ on $S_c$ and $S_t$, and the unsupervised alignment loss between $M_c$ and $M_t$ on $S_u$, respectively.
Since they all have the corresponding supervision information, all the three terms can be directly optimized.
Under the triangle inequality, term $(b.1)$ can be seen as the difference between both $S_c$ and $S'_{c}$ when $S_c$'s corresponding feedback $S'_{c}$ in $R^t$ is known.
Therefore, term $(b.1)$ is a constant that can be used to estimate the degree of difference between the two policies, and is usually small since the system-induced biases do not have an excessive effect on the user's true preference.
The term $(e.1)$ is the supervised loss of $M_t$ itself on $S_t$, and thus can also be directly optimized.
Under the separability, term $(b.2)$ and term $(c)$ jointly adjust $M_c$'s trade-off in the supervised loss on $S_c$. 
Since we do not have the true feedback labels of $R_t$ on the specific user-item pair index set $I_{S_c}$, we cannot directly optimize the term $(b.2)$. 
Fortunately, our experiments show that our method still has a significant advantage even in its absence, and we leave its further processing as future work.
Similarly, term $(e.2)$ and term $(a)$ jointly adjust $M_c$'s trade-off in the supervised loss on $S_t$. 
Since the prediction error of $M_t$ on the specific user-item pair index set $I_{S_t}$ is available, term $(e.2)$ can also be directly optimized.
In short, no matter which generalization error bound is satisfied by the adopted loss function, we can improve the unbiased performance of the recommendation model by simultaneously minimizing the terms $(a)$, $(c)$, $(d)$, and $(e)$ in the generalization error bound.
Note that the last two terms in the generalization error bound as shown in Eq.(\ref{equ:triangle}) are the error terms that arise when we use the easy-to-solve term $(e)$ in Eq.(\ref{equ:triangle}) to approximate the fourth difficult-to-solve term in Proposition~\ref{prop:triangle}.
Their values depend on the confidence of this approximation process and are independent of the model.
In particular, we can find that as the size of a randomized dataset gradually increases, the values of these error terms gradually decrease, which means that the approximation operation is more reliable.
This is expected, that when a randomized dataset is large, the training of the model can benefit more from more reliable unbiased information.
The last two terms of another generalization error bound shown in Eq.(\ref{equ:separability}) have similar properties.

\subsection{Debiasing Approximate Upper Bound with A Randomized Dataset}
Based on the analysis for each term of the generalization error bound in Sec.~\ref{subsec:analysis}, we propose a novel method called debiasing approximate upper bound with a randomized dataset (DUB), which aims to directly optimize the upper bound of the unbiased ideal loss function.
Note that we use the term ``approximate upper bound'' to distinguish it from the term ``upper bound'' since our DUB considers the terms in Eq.(\ref{equ:triangle}) (or Eq.(\ref{equ:separability})) that can be directly optimized but not all the terms.
Specifically, depending on the types of the loss functions used, we have two types of objective functions to be optimized. 
When the used loss function satisfies the triangular inequality, the optimization goal is shown in Eq.(\ref{equ:bnu_1}), which is to minimize a proxy of the upper bound shown in Eq.(\ref{equ:triangle}).
\begin{equation}\label{equ:bnu_1}
    \begin{aligned}
    \mathop{\min}\limits_{\mathcal{W}_c,\mathcal{W}_t} &\underbrace{\mathcal{L}^{S_t}\left(R^t,\hat{R}^c\right)}_{(a)} + \underbrace{\mathcal{L}^{S_c}\left(R^c,\hat{R}^c\right)}_{(c)} + \underbrace{\mathcal{L}_{|S_t|}^{S_t}\left(R^t,\hat{R}^t\right)}_{(e.1)} + \gamma\underbrace{\mathcal{L}^{S_u}\left(\hat{R}^t,\hat{R}^c\right)}_{(d)} +  \lambda_c \rm{Reg}\left(\mathcal{W}_c\right) + \lambda_t \rm{Reg}\left(\mathcal{W}_t\right),
    \end{aligned}
\end{equation}
where $\gamma$ is the weight parameter of $\mathcal{L}^{S_u}\left(\hat{R}^t,\hat{R}^c\right)$, and $\mathcal{W}_c$ and $\mathcal{W}_t$ denote the parameters of $M_c$ and $M_t$, respectively. 
Note that $\rm{Reg}\left(\cdot\right)$ is the regularization term, and $\lambda_c$ and $\lambda_t$ are the parameters of the regularization. 
Recall from the analysis in Sec.~\ref{subsec:analysis} that all the terms that can be directly optimized in the generalization error bound as shown in Eq.(\ref{equ:triangle}) include the terms $(a)$, $(c)$, $(d)$, and $(e.1)$.
This corresponds to each optimization term in Eq.(\ref{equ:bnu_1}).
Note that since the unsupervised loss of $M_t$ and $M_c$ on $S_u$ may contain too much noise when the size of a randomized dataset $S_t$ is small, we introduce a weight parameter $\gamma$ to control its influence.
In addition, for the stability of model training, we also additionally include two regularization terms for the model parameters.
An intuitive explanation of Eq.(\ref{equ:bnu_1}) is to use a non-randomized dataset $S_c$ and a randomized dataset $S_t$ for the trade-off learning of $M_c$, and to further provide the unbiased information for $M_c$ through the imputation labels provided by $M_t$.
Therefore, our DUB can be viewed as a combination of sample-based debiasing distillation and label-based debiasing distillation defined in~\cite{liu2020general}.

When the used loss function satisfies the separability, the optimization problem is shown in Eq.(\ref{equ:bnu_2}), which is to minimize a proxy of the upper bound shown in Eq.(\ref{equ:separability}).
\begin{equation}\label{equ:bnu_2}
    \begin{aligned}
    \mathop{\min}\limits_{\mathcal{W}_c} &\underbrace{\mathcal{L}^{S_t}\left(R^t,\hat{R}^c\right)}_{(a)} + \underbrace{\mathcal{L}^{S_c}\left(R^c,\hat{R}^c\right)}_{(c)} + \underbrace{\mathcal{L}_{|S_t|}^{S_t}\left(R^t-\hat{R}^t,\hat{R}^c\right)}_{(e.2)} +
    \gamma\underbrace{\mathcal{L}^{S_u}\left(\hat{R}^t,\hat{R}^c\right)}_{(d)} +  \lambda_c \rm{Reg}\left(\mathcal{W}_c\right).
    \end{aligned}
\end{equation}
Similarly, based on the analysis in Sec.~\ref{subsec:analysis}, all the terms that can be directly optimized in the generalization error bound shown in Eq.(\ref{equ:separability}) include the terms $(a)$, $(c)$, $(d)$, and $(e.2)$.
This corresponds to each optimization term in Eq.(\ref{equ:bnu_2}).
For the same reason, we also additionally introduce a weight parameter $\gamma$ and a regularization term for the model parameters.
Note that no supervised loss related to $M_t$ is included in Eq.(\ref{equ:bnu_2}), so we only introduce a regularization term for $M_c$.
An intuitive explanation of Eq.(\ref{equ:bnu_2}) is similar to Eq.(\ref{equ:bnu_1}), except that Eq.(\ref{equ:bnu_2}) additionally includes an optimization term (i.e., term $(e.2)$) to enhance $M_c$'s learning of $S_t$.
This can make the model more robust when the relative unbiasedness of a randomized dataset is not high, such as being affected by some business rules.
Regardless of whether Eq.(\ref{equ:bnu_1}) or Eq.(\ref{equ:bnu_2}) is used, the proposed method includes all the terms that can be directly optimized as analyzed in Sec.~\ref{subsec:analysis}. 
Obviously, our method is a more sufficient optimization of the upper bounds, which is expected to further improve the performance. 

However, in real applications, we observe an implied limitation of our method due to the large difference in the number of non-uniform data $S_c$ and the uniform data $S_t$. 
Since the scale of $S_c$ is usually much larger than that of $S_t$, this will lead to the inconsistency of training difficulty between $M_c$ and $M_t$, i.e., $M_t$ will converge faster.
This asynchrony will have an undesirable effect on the prediction alignment term, i.e., $\mathcal{L}^{S_u}\left(\hat{R}^t,\hat{R}^c\right)$. 
Finally, the overall training is unstable. 
To alleviate this problem, we first pre-train $M_c$ and $M_t$. 
Subsequently, we refine the pre-trained models again according to the above loss function. 
The pseudo code of DUB is shown in Algorithm~\ref{alg:dub}.

Note that similar to most existing debiasing methods, our DUB does not depend on a specific model architecture when deploying or applying it in practice.
Specifically, the process of integrating our DUB into any recommendation model is as follows:
1) after collecting a non-randomized dataset $S_c$ and a randomized dataset $S_t$, we pre-train a recommendation model $M_c$ and an auxiliary model $M_t$ based on a traditional optimization objective function and an arbitrary recommendation model, respectively (lines 1 and 2 of Algorithm~\ref{alg:dub});
and 2) in the model refinement stage, we only need to modify the optimization objective function of these models to that of DUB in the training stage, i.e., according to the type of loss function used, we choose Eq.(\ref{equ:bnu_1}) or Eq.(\ref{equ:bnu_2}) as the new objective function (lines 4 to 6 of Algorithm~\ref{alg:dub}).

\begin{algorithm}[htbp]
	\caption{Debiased Upper Bound with A Randomized Dataset (DUB)}
	\begin{algorithmic}[1]
		\REQUIRE A non-randomized dataset $S_c$ and a randomized dataset $S_t$.
		\STATE Train a pre-trained recommendation model $M_c$ based on a backbone model on $S_c$.
		\STATE Train a pre-trained auxiliary model $M_t$ based on a backbone model on $S_t$.
		\STATE \textbf{repeat}
		\STATE An auxiliary set $S_a$ with the same size as the training sample is randomly sampled from the unobserved feedback $S_u$;
		\STATE Based on $S_c$, $S_t$ and $S_a$, use the pre-trained $M_c$ and $M_t$ to calculate each loss term in Eq.(\ref{equ:bnu_1}) or Eq.(\ref{equ:bnu_2}) (according to the conditions satisfied by the adopted loss function);
		\STATE Update the parameters of the recommendation model $M_c$.
		\STATE \textbf{until} convergence
	\end{algorithmic}
\label{alg:dub}
\end{algorithm}

\section{Analysis of Existing Methods}\label{sec:existing_baselines}
In this section, we will introduce and analyze some existing methods. 
In particular, different from the proposed method, we show that these methods only optimize some terms in the generalization error bounds of the unbiased ideal loss function, or optimize some weak proxy of these terms, i.e., an insufficient optimization of the generalization error bound.
This means that these methods may only converge to a sub-optimal solution. 
Note that an insufficient optimization for the generalization error bound is different from a more compact generalization error bound.
The former means that the model only considers some optimization items and ignores the constraints on some optimization items during the training process.
This may lead to the fact that although some optimization terms are gradually minimized, the generalization error bound may be unchanged, and even grow in reverse, due to the gradual increase in the loss of the neglected optimization terms.
The latter means that it is closer to the ideal optimization objective function than the other generalization error bounds.

\subsection{Causal Embeddings}
Causal Embeddings (CausE)~\cite{schnabel2016recommendations} is a pioneering work in counterfactual recommendation. 
By introducing causal inference into the representation learning of recommendation, CausE is implemented in a multi-task learning framework, including a treatment task loss (i.e., $M_c$'s own supervised loss), a control task loss (i.e., $M_t$'s own supervised loss), and a regularizer between tasks (i.e., the parameter alignment terms of $M_c$ and $M_t$). 
In particular, the loss function of CausE can be written as follows,
\begin{equation}\label{equ:cause}
    \begin{aligned}
    \mathop{\min}\limits_{\mathcal{W}_c,\mathcal{W}_t} &\underbrace{\mathcal{L}^{S_c}\left(R^c,\hat{R}^c\right)}_{(c)} +\underbrace{\mathcal{L}^{S_t}\left(R^t,\hat{R}^t\right)}_{(e.1)}+\lambda_c \rm{Reg}\left(\mathcal{W}_c\right)+ 
    \lambda_t \rm{Reg}\left(\mathcal{W}_t\right) +\gamma_{tc}^{CausE}\underbrace{\left\|\mathcal{W}_t-\mathcal{W}_c\right\|_F}_{(\mathit{d})},
    \end{aligned}
\end{equation}
where $\gamma_{tc}^{CausE}$ is the weight parameter of the alignment term between $M_c$ and $M_t$.

By comparing Eq.(\ref{equ:cause}) with Theory~\ref{theorem:triangle}, the objective function of CausE can be regarded as a combination of term $(c)$, term $(e.1)$ and a proxy of term $(d)$ (i.e.,  $\left\|\mathcal{W}_t-\mathcal{W}_c\right\|_F$). 
Similarly, in comparison with Theory~\ref{theorem:separability}, it can be regarded as a combination of term $(c)$ and a proxy of term $(d)$. 
This means that CausE is an insufficient optimization of the generalization error bound. 
In addition, we find that the parameter alignment term may not be a reasonable proxy for the term $(d)$: 
1) The parameter alignment term restricts the parameters of $M_c$ and $M_t$ to have the same dimension. However, in view of the difference in data scale between $S_c$ and $S_t$, this constraint may be too strong. 
2) The alignment of the parameters will cause difficulty in training in case of high dimensions and multi-layer networks. 
The lack of optimization for terms $(a)$ and $(e.2)$ will also result in CuasE not being able to make $M_c$ fully benefit from $S_t$ during training, especially when $S_t$ has a particularly small scale.

\subsection{Bridge Strategy}
Recently, Liu et al. explain and resolve counterfactual recommendation from the perspective of knowledge distillation~\cite{liu2020general}. 
They propose a general knowledge distillation framework for counterfactual recommendation, and list some practical solutions as examples. 
The Bridge strategy is one of these solutions with the best performance, which also best matches our focus. 
The Bridge strategy first ensures the supervised loss of $M_c$ and $M_t$. 
In addition, an auxiliary set $S_a$ is randomly sampled from $\mathcal{D}$ in each iteration, and the predictions of $M_c$ and $M_t$ in $S_a$ are constrained to be close. 
Note that most of $S_a$ belong to $S_u$ because of the data sparsity in recommender systems. In particular, the loss function of the Bridge strategy can be rewritten as follows,
\begin{equation}\label{equ:bridge}
    \begin{aligned}
    \mathop{\min}\limits_{\mathcal{W}_c,\mathcal{W}_t} &\underbrace{\mathcal{L}^{S_c}\left(R^c,\hat{R}^c\right)}_{(c)} + \underbrace{\mathcal{L}^{S_t}\left(R^t,\hat{R}^t\right)}_{(e.1)} + \gamma\mathcal{L}^{S_a}\underbrace{\left(\hat{R}^t,\hat{R}^c\right)}_{(d)} + \lambda_c \rm{Reg}\left(\mathcal{W}_c\right) + \lambda_t \rm{Reg}\left(\mathcal{W}_t\right).
    \end{aligned}
\end{equation}
By comparing Eq.(\ref{equ:bridge}) with Theory~\ref{theorem:triangle}, the objective function of Bridge can be regarded as a combination of terms $(c)$, $(e.1)$ and $(d)$. 
Similarly, in comparison with Theory~\ref{theorem:separability}, it can be regarded as a combination of terms $(c)$ and $(d)$. 
This means that the Bridge strategy is also an insufficient optimization of the generalization error bound. 
But it directly optimizes term $(d)$ instead of using a weak proxy, and thus achieves a better performance in the experiments~\cite{liu2020general}.
Similarly, the lack of optimization for terms $(a)$ and $(e.2)$ can also cause Bridge to fail to make $M_c$ fully benefit from $S_t$ in some cases.

\subsection{Remarks}
Note that our discussion does not include another recent method AutoDebias~\cite{chen2021autodebias}, where meta-learning is introduced into a doubly robust (DR) framework to learn better unbiased parameters. 
On one hand, it can be seen as an improvement in the training process rather than in the loss function, which is different from our DUB as well as the existing methods mentioned above.
On the other hand, the DR framework is also a representative method in another route without a randomized dataset~\cite{wang2019doubly,guo2021enhanced}, i.e., a randomized dataset is not necessary.
Therefore, it is difficult to put it into a specific category.
In addition, some theoretical insights on debiased recommendation are also provided in~\cite{chen2021autodebias}, which is however quite different from our DUB.
In particular, they aim to analyze the theoretical generalization error bound of AutoDebias, while we directly optimize a proxy of the upper bound derived from the unbiased ideal loss function in Eq.(\ref{equ:st_ideal_estimator}).

\section{Empirical Evaluation}\label{sec:empirical}
In this section, we conduct experiments with the aim of answering the following five key questions. 
Note that the source codes and results are available at \url{https://github.com/dgliu/TOIS_DUB}.
\begin{itemize}[leftmargin=*]
    \item RQ1: How does the proposed method perform against the baselines in an unbiased evaluation?
    \item RQ2: What is the role of the additional terms in the loss function of the proposed method (i.e., the ablation studies of our DUB)?
    \item RQ3: What impact does the proposed method have on the item distribution of the recommendation lists?
    \item RQ4: How do some key factors affect the performance of the proposed method?
    \item RQ5: How does the proposed method perform against the baselines in a general biased evaluation?
\end{itemize}

\subsection{Experimental Setup}
\subsubsection{Datasets.}\label{subsubsec:datasets} To evaluate the performance of the model in an ideal unbiased scenario, we need to use a dataset containing a randomized subset. 
We thus use the following two datasets in the experiments, where the statistics are shown in Table~\ref{tab:datasets}.
\begin{itemize}[leftmargin=*]
    \item \textbf{Yahoo! R3}~\cite{marlin2009collaborative}: This is the most commonly adopted standard dataset in previous works, including a user subset and a random subset. 
    The former can be regarded as being collected under a stochastic recommendation policy, while the latter corresponds to a uniform policy. 
    We binarize the ratings via a threshold $\epsilon=3$, where a rating $>\epsilon$ is considered as a positive feedback (i.e., $R_{ij}=1$); and otherwise, it is a negative feedback (i.e., $R_{ij}=0$). 
    The user subset is used as a training set in a biased environment ($S_c$). 
    For the random subset, we randomly split the user-item interactions into three subsets, including 10\% for training in an unbiased environment ($S_t$), 10\% for validation to tune the hyper-parameters ($S_{va}$), and the rest 80\% for test ($S_{te}$). 
    \item \textbf{Product}: This is a large-scale dataset for CTR prediction, which includes two weeks of users' click records from a real-world advertising system. 
    The dataset contains two subsets: a subset ($S_c$) logged by several traditional ranking policies and a subset ($S_t$) logged by a uniform policy $\pi_t$. 
    To remove the effect of the position bias in our experiments, we filter out the samples at positions 1 and 2. 
    The dataset covers 217 displayed ads and more than two million users. 
    To get the training set, validation set and test set from the uniform subset, we randomly split the $S_t$ subset using the same proportions as that for Yahoo! R3. 
\end{itemize}

\begin{table}[htbp]
\caption{Statistics of the datasets. P/N denotes the ratio between the numbers of positive and negative feedback.}
\centering
\scalebox{0.9}{
\begin{tabular}{c|cc|cc}
\specialrule{0.1em}{3pt}{3pt}
 & \multicolumn{2}{c|}{\textbf{Yahoo! R3}} & \multicolumn{2}{c}{\textbf{Product}} \\
 & \#Feedback & P/N& \#Feedback & P/N\\
\specialrule{0.05em}{3pt}{3pt}
$S_c$ &311,704 &67.02\% &4,798,776 &12.97\%\\
$S_t$ &5,400 &9.05\% &34,755 &0.99\%\\
$S_{va}$ &5,400 &9.31\% &34,755 &0.75\%\\
$S_{te}$ &43,200 &9.76\% &278,043 &0.88\%\\
\specialrule{0.1em}{3pt}{3pt}
\end{tabular}}
\label{tab:datasets}
\end{table}

\subsubsection{Backbones.} The debiasing methods are usually model agnostic and are integrated into some backbone models. 
To comprehensively evaluate the generalization ability, we use two representative shallow and deep models as the backbone models in the experiments, i.e., matrix factorization (MF)~\cite{koren2009matrix} and neural collaborative filtering (NCF)~\cite{he2017neural}. 
Similar settings can be found in previous works~\cite{schnabel2016recommendations,bonner2018causal,saito2020asymmetric,wang2020information}.

\subsubsection{Baselines.} For the basic model, it can be regarded as three variants according to the different data sources used, i.e., training only with a non-randomized dataset $S_c$, training only with a randomized dataset $S_t$, and training with both data (i.e., $S_c \cup S_t$). 
We call the latter two variants \textit{Unif} and \textit{Combine} in the experiments. 
For debiased recommendation models, we choose the representative methods among the three lines summarized in Sec.~~\ref{intro}. 
For the first line, the inverse propensity score (IPS)~\cite{schnabel2016recommendations} is one of the most classic methods, which thus also serves as one of our baselines. 
We adopt the na\"{\i}ve Bayes estimator in~\cite{schnabel2016recommendations} to estimate the propensity score. 
For the second line, a recent method AutoDebias is introduced, in which the information of a randomized dataset is used more effectively by combining meta-learning strategies in a doubly robust framework~\cite{chen2021autodebias}.
For the third line, as described in Sec.~\ref{sec:existing_baselines}, CausE~\cite{bonner2018causal} and Bridge~\cite{liu2020general} are two important baselines because they are the state-of-the-art methods that best match our focus. 

\subsubsection{Evaluation Metrics.} We employ four evaluation metrics that are widely used in recommender systems, including precision (P@K), recall (R@K), the area under the ROC curve (AUC)  and normalized discounted cumulative gain (nDCG). We choose AUC as our main evaluation metric because it is one of the most important metrics in industry and previous works on debiasing. We report the results with $K$ set to 5 and 10. The candidate items to be recommended for a user are from the set of items that have not been interacted by the user.

\subsubsection{Implementation Details.}\label{subsubsec:imp} All the methods are implemented on TensorFlow 1.2~\cite{abadi2016tensorflow}, except \textit{AutoDebias} referring to its official PyTorch~\cite{paszke2019pytorch} version. We use the Adam~\cite{kingma2014adam} optimizer and the cross-entropy loss in the experiments, i.e., we choose Eq.(\ref{equ:bnu_2}) as the optimization objective of the model. The learning rate is fixed as $1e^{-3}$. By evaluating the AUC on the validation data $S_{va}$, we perform grid search to tune the hyper-parameters for the candidate methods. 
To avoid over-fitting, we adopt an early stopping mechanism with the patience set to 5 times.
The range of the values of the hyper-parameters are shown in Table~\ref{tab:search_ranges}.
\begin{table}[htbp]
\caption{Hyper-parameters tuned in the experiments.}
\centering
\scalebox{0.9}{
\begin{tabular}{ccc}
\specialrule{0.1em}{3pt}{3pt}
\textbf{Name} & \textbf{Range} & \textbf{Functionality}\\
\specialrule{0.05em}{3pt}{3pt}
$rank$ & $\left\{50,100,200\right\}$ & Embedded dimension \\
$\lambda$ &$\left\{1e^{-5},1e^{-4}\cdots1e^{-1}\right\}$ & Regularization \\
$\gamma$ &$\left\{1e^{-5},1e^{-4}\cdots1e^{-1}\right\}$ & Loss weighting \\
\specialrule{0.1em}{3pt}{3pt}
\end{tabular}}
\label{tab:search_ranges}
\vspace{-5pt}
\end{table}

\subsection{RQ1: Comparison Results of Unbiased Evaluation}\label{exp:results_unbiased}
We report the comparison results of the unbiased evaluation in Table~\ref{tab:comparison_result_1} and Table~\ref{tab:comparison_result_2}. 
For the Yahoo! R3 dataset, as shown in Table~\ref{tab:comparison_result_1}, the proposed method outperforms all baselines in most cases except on P@5 and R@5 when NCF is used as the backbone model.
Specifically, we have the following observations: 1) The baselines based on the use of a randomized dataset usually have a better performance than the basic model, but may suffer from a performance bottleneck in some cases. 2) The performance of the baseline AutoDebias depends on the backbone model used, which may be because the designed meta-learning strategy is mainly for low-rank models. 3) On the contrary, our DUB is relatively stable for different backbone models.
For the Product dataset, as shown in Table~\ref{tab:comparison_result_2}, the proposed method consistently outperforms all the baselines on AUC, and maintains advantages on other metrics in most cases. We can get similar observations as that on Yahoo! R3. 
Note that since the baseline AutoDebias has a prediction step for all the unobserved samples, it requires far more memory than that of a single GPU (e.g., 32G) and a specific parallelization. 
This weakens its scalability, and we do not report its results.
In general, our DUB is relatively stable for datasets of different sizes.

\begin{table}[htbp]
\caption{Comparison results of unbiased testing on Yahoo! R3, where the best results are marked in bold. AUC is the main evaluation metric. Note that $^{*}$ indicates a significance level $p\leq 0.05$ based on two sample t-test between the best and second best results.}
\centering
\scalebox{0.9}{
\begin{tabular}{c|cccccc}
\specialrule{0.1em}{3pt}{3pt}
Method& AUC & nDCG & P@5 & P@10 & R@5 & R@10\\  
\specialrule{0.05em}{3pt}{3pt}
MF & 0.7282 & 0.0434 & 0.0059 & 0.0051 & 0.0207 & 0.0332 \\
Unif-MF & 0.5625 & 0.0291 & 0.0049 & 0.0041 & 0.0135 & 0.0245 \\
Combine-MF  & 0.7357 & 0.0489 & 0.0073 & 0.0061 & 0.0243 & 0.0401 \\
\specialrule{0.05em}{3pt}{3pt}
IPS-MF & 0.7300 & 0.0407 & 0.0052 & 0.0054 & 0.0171 & 0.0344 \\
AutoDebias-MF & 0.7502 & 0.0691 & 0.0119 & 0.0104 & 0.0403 & 0.0683 \\
CausE-MF & 0.7285 & 0.0445 & 0.0059 & 0.0058 & 0.0192 & 0.0372 \\
Bridge-MF & 0.7376 & 0.0557 & 0.0099 & 0.0076 & 0.0308 & 0.0478 \\
\specialrule{0.05em}{3pt}{3pt}
DUB-MF & $\textbf{0.7578}^{*}$ & \textbf{0.0727} & \textbf{0.0128} & \textbf{0.0112} & \textbf{0.0438} & \textbf{0.0770}\\
\specialrule{0.1em}{3pt}{3pt}

\specialrule{0.1em}{3pt}{3pt}
NCF & 0.7245 & 0.0279 & 0.0029 & 0.0031 & 0.0089 & 0.0199 \\
Unif-NCF & 0.6050 & 0.0275 & 0.0043 & 0.0037 & 0.0113 & 0.0204 \\
Combine-NCF  & 0.7268 & 0.0327 & 0.0032 & 0.0033 & 0.0092 & 0.0189 \\
\specialrule{0.05em}{3pt}{3pt}
IPS-NCF & 0.7273 & 0.0304 & 0.0036 & 0.0031 & 0.0111 & 0.0210 \\
AutoDebias-NCF & 0.7140 & 0.0385 & 0.0052 & 0.0047 & 0.0188 & 0.0333 \\
CausE-NCF & 0.7284 & 0.0287 & 0.0029 & 0.0033 & 0.0089 & 0.0210 \\
Bridge-NCF & 0.7367 & 0.0439 & \textbf{0.0056} & 0.0056 & \textbf{0.0192} & 0.0371 \\
\specialrule{0.05em}{3pt}{3pt}
DUB-NCF & $\textbf{0.7421}^{*}$ & \textbf{0.0491} & 0.0051 & \textbf{0.0058} & 0.0164 & \textbf{0.0390} \\
\specialrule{0.1em}{3pt}{3pt}
\end{tabular}}
\label{tab:comparison_result_1}
\end{table}

\begin{table}[htbp]
\caption{Comparison results of unbiased testing on Product, where the best results are marked in bold. AUC is the main evaluation metric. Note that $^{*}$ indicates a significance level $p\leq 0.05$ based on two sample t-test between the best and second best results.}
\centering
\scalebox{0.9}{
\begin{tabular}{c|cccccc}
\specialrule{0.1em}{3pt}{3pt}
Method& AUC & nDCG & P@5 & P@10 & R@5 & R@10\\  
\specialrule{0.05em}{3pt}{3pt}
MF & 0.7115 & 0.0434 & 0.0105 & 0.0103 & 0.0518 & 0.1017 \\
Unif-MF & 0.6372 & 0.0604 & 0.0148 & 0.0135 & 0.0737 & 0.1332 \\
Combine-MF  & 0.7145 & 0.0526 & 0.0121 & 0.0113 & 0.0601 & 0.1111 \\
\specialrule{0.05em}{3pt}{3pt}
IPS-MF & 0.7274 & 0.0484 & 0.0115 & 0.0114 & 0.0568 & 0.1114 \\
AutoDebias-MF & - & - & - & - & - & - \\
CausE-MF & 0.7158 & 0.0470 & 0.0107 & 0.0114 & 0.0529 & 0.1119 \\
Bridge-MF  & 0.7069 & 0.0438 & 0.0107 & 0.0104 & 0.0529 & 0.1022 \\
\specialrule{0.05em}{3pt}{3pt}
DUB-MF & $\textbf{0.7374}^{*}$ & \textbf{0.0729} & \textbf{0.0158} & \textbf{0.0155} & \textbf{0.0787} & \textbf{0.1537} \\
\specialrule{0.1em}{3pt}{3pt}

\specialrule{0.1em}{3pt}{3pt}
NCF & 0.7293 & 0.0616 & 0.0152 & 0.0131 & 0.0753 & 0.1299 \\
Unif-NCF & 0.6240 & 0.0557 & 0.0131 & 0.0132 & 0.0651 & 0.1307 \\
Combine-NCF  & 0.7301 & 0.0674 & 0.0155 & 0.0142 & 0.0773 & \textbf{0.1410} \\
\specialrule{0.05em}{3pt}{3pt}
IPS-NCF & 0.7328 & 0.0616 & 0.0155 & 0.0126 & 0.0773 & 0.1249 \\
AutoDebias-NCF & - & - & - & - & - & - \\
CausE-NCF & 0.7351 & 0.0623 & 0.0158 & 0.0125 & 0.0789 & 0.1235 \\
Bridge-NCF  & 0.7149 & 0.0628 & 0.0145 & 0.0126 & 0.0723 & 0.1255 \\
\specialrule{0.05em}{3pt}{3pt}
DUB-NCF & $\textbf{0.7382}^{*}$ & \textbf{0.0686} & \textbf{0.0165} & \textbf{0.0149} & \textbf{0.0851} & 0.1380 \\
\specialrule{0.1em}{3pt}{3pt}
\multicolumn{7}{c}{\tabincell{l}{$\bullet$ Note: the placeholder `-' means that the result is not reported because the\\ $\ \ $ memory space required by this method exceeds that of the GPU used.}}
\end{tabular}}
\label{tab:comparison_result_2}
\end{table}

\subsection{RQ2: Results of Ablation Studies}\label{exp:results_ablation}
As described in Sec.~\ref{sec:method}, the proposed method further improves the performance by sufficiently optimizing the upper bound of the unbiased ideal loss function. 
A key question is what the role of the additional optimization terms is in our method.
To answer this question, we conduct ablation studies of the proposed method by removing certain terms.
The results are shown in Table~\ref{tab:ablation_1} and Table~\ref{tab:ablation_2}.
Note that after removing terms $(a)$ and $(e)$, our DUB is equivalent to the Bridge strategy, so we do not remove more terms in the experiments.
We can see that removing any term will hurt the performance in most cases, and removing more terms results in worse performance.
There are some unexpected cases in Table~\ref{tab:ablation_1}, i.e., when K takes a small value, the full version with NCF as the backbone model has a slight disadvantage on a few metrics. This may be due to the noise caused by only considering AUC as the evaluation metric in parameter tuning.
In general, all terms in the proposed method can synergistically produce the greatest gain.
\begin{table}[htbp]
\caption{Results of the ablation studies on Yahoo! R3, where the best results are marked in bold. AUC is the main evaluation metric.}
\centering
\scalebox{0.9}{
\begin{tabular}{l|cccccc}
\specialrule{0.1em}{3pt}{3pt}
Method& AUC & nDCG & P@5 & P@10 & R@5 & R@10\\ 
\specialrule{0.05em}{3pt}{3pt}
DUB-MF & \textbf{0.7578} & \textbf{0.0727} & \textbf{0.0128} & \textbf{0.0112} & \textbf{0.0438} & \textbf{0.0770} \\
\specialrule{0.05em}{3pt}{3pt}
w/o term $\left(e.2\right)$ & 0.7500 & 0.0702 & 0.0113 & 0.0108 & 0.0377 & 0.0744 \\
\specialrule{0.05em}{3pt}{3pt}
w/o terms $\left(a\right) \& \left(e.2\right)$ & 0.7376 & 0.0557 & 0.0099 & 0.0076 & 0.0308 & 0.0478 \\
\specialrule{0.1em}{3pt}{3pt}
\specialrule{0.1em}{3pt}{3pt}
DUB-NCF & \textbf{0.7421} & \textbf{0.0491} & 0.0051 & \textbf{0.0058} & 0.0164 & \textbf{0.0390} \\
\specialrule{0.05em}{3pt}{3pt}
w/o term $\left(e.2\right)$  & 0.7386 & 0.0438 & 0.0050 & 0.0051 & 0.0165 & 0.0320 \\
\specialrule{0.05em}{3pt}{3pt}
w/o terms $\left(a\right) \& \left(e.2\right)$ & 0.7367 & 0.0439 & \textbf{0.0056} & 0.0056 & \textbf{0.0192} & 0.0371 \\
\specialrule{0.1em}{3pt}{3pt}
\end{tabular}}
\label{tab:ablation_1}
\end{table}

\begin{table}[htbp]
\caption{Results of the ablation studies on Product, where the best results are marked in bold. AUC is the main evaluation metric.}
\centering
\scalebox{0.9}{
\begin{tabular}{l|cccccc}
\specialrule{0.1em}{3pt}{3pt}
Method& AUC & nDCG & P@5 & P@10 & R@5 & R@10\\  
\specialrule{0.05em}{3pt}{3pt}
DUB-MF & \textbf{0.7374} & \textbf{0.0729} & \textbf{0.0158} & \textbf{0.0155} & \textbf{0.0787} & \textbf{0.1537} \\
\specialrule{0.05em}{3pt}{3pt}
w/o term $\left(e.2\right)$ & 0.7091 & 0.0453 & 0.0115 & 0.0105 & 0.0571 & 0.1039 \\
\specialrule{0.05em}{3pt}{3pt}
w/o terms $\left(a\right) \& \left(e.2\right)$ & 0.7069 & 0.0438 & 0.0107 & 0.0104 & 0.0529 & 0.1022 \\
\specialrule{0.1em}{3pt}{3pt}
\specialrule{0.1em}{3pt}{3pt}
DUB-NCF & \textbf{0.7382} & \textbf{0.0686} & \textbf{0.0165} & \textbf{0.0149} & \textbf{0.0851} & \textbf{0.1380} \\
\specialrule{0.05em}{3pt}{3pt}
w/o term $\left(e.2\right)$ & 0.7284 & 0.0648 & 0.0162 & 0.0132 & 0.0806 & 0.1313 \\
\specialrule{0.05em}{3pt}{3pt}
w/o terms $\left(a\right) \& \left(e.2\right)$ & 0.7149 & 0.0628 & 0.0145 & 0.0126 & 0.0723 & 0.1255 \\
\specialrule{0.1em}{3pt}{3pt}
\end{tabular}}
\label{tab:ablation_2}
\end{table}

\subsection{RQ3: Item Distribution of the Recommendation Lists}\label{exp:results_item}
\begin{figure}
\centering
    \subfigure[]{
    \label{fig:ratio_random}
    \includegraphics[width=0.48\columnwidth]{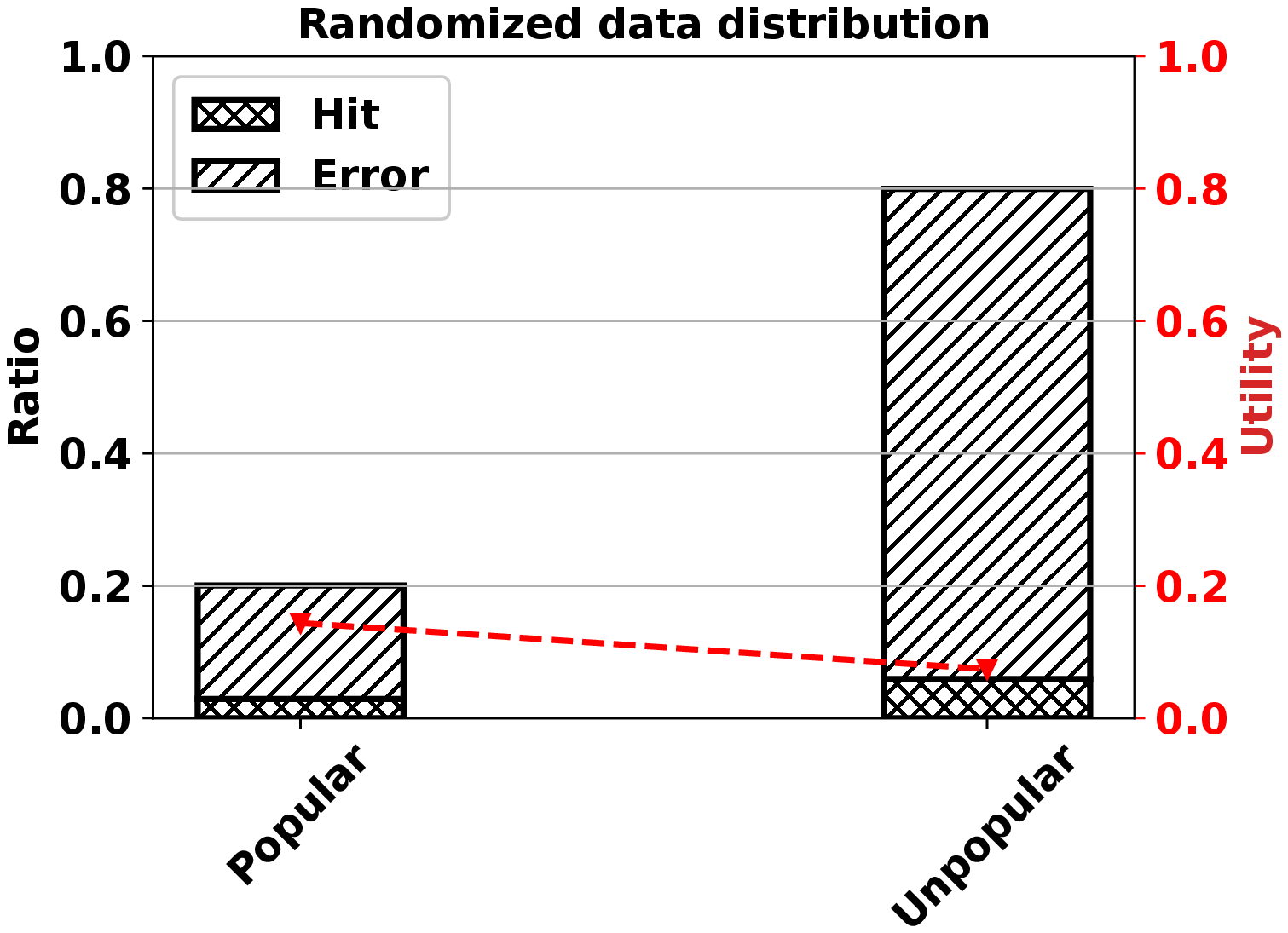}}\\
    \subfigure[]{
    \label{fig:ratio_pos}
    \includegraphics[width=0.48\columnwidth]{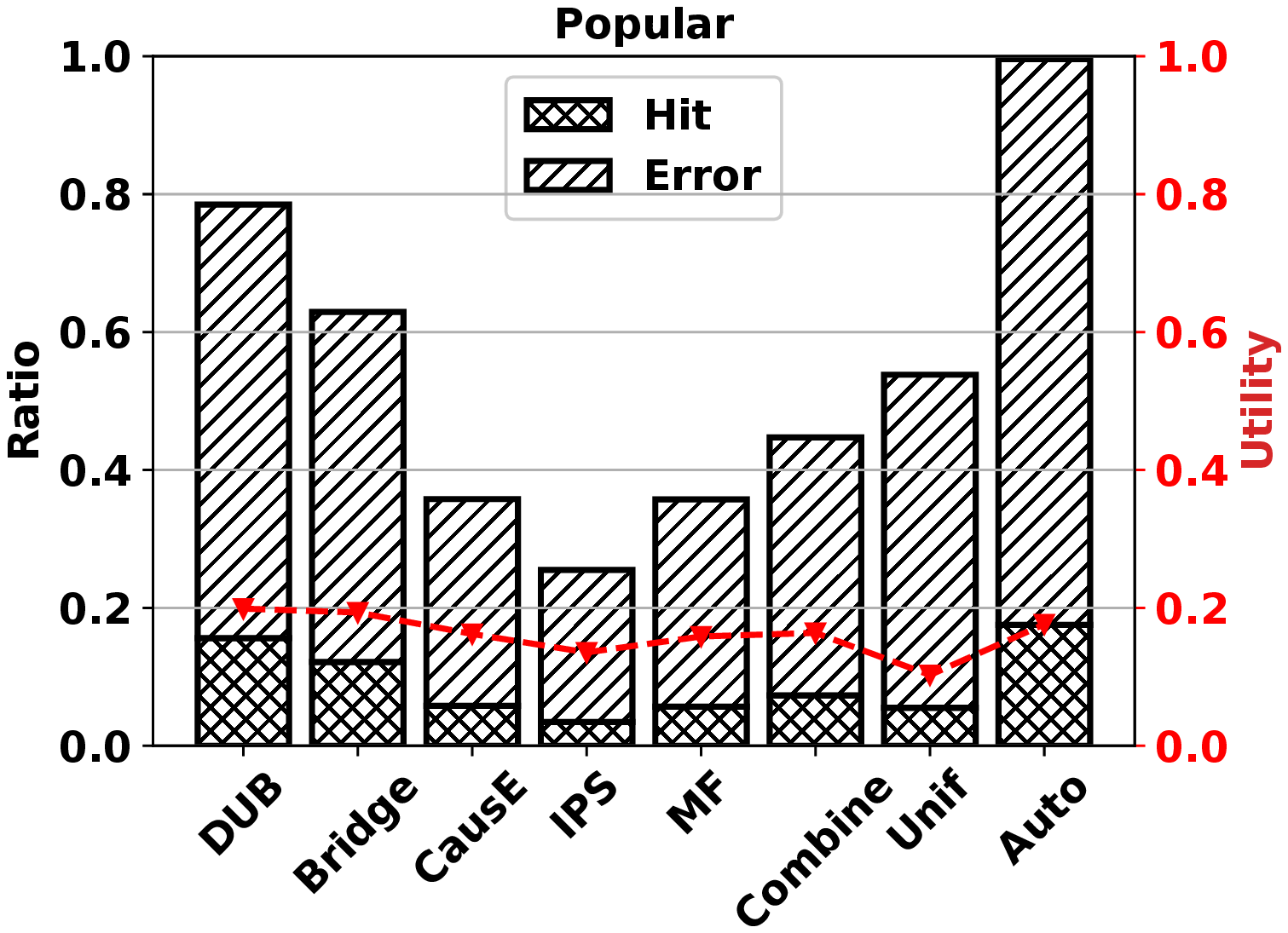}}
    \subfigure[]{
    \label{fig:ratio_unpos}
    \includegraphics[width=0.48\columnwidth]{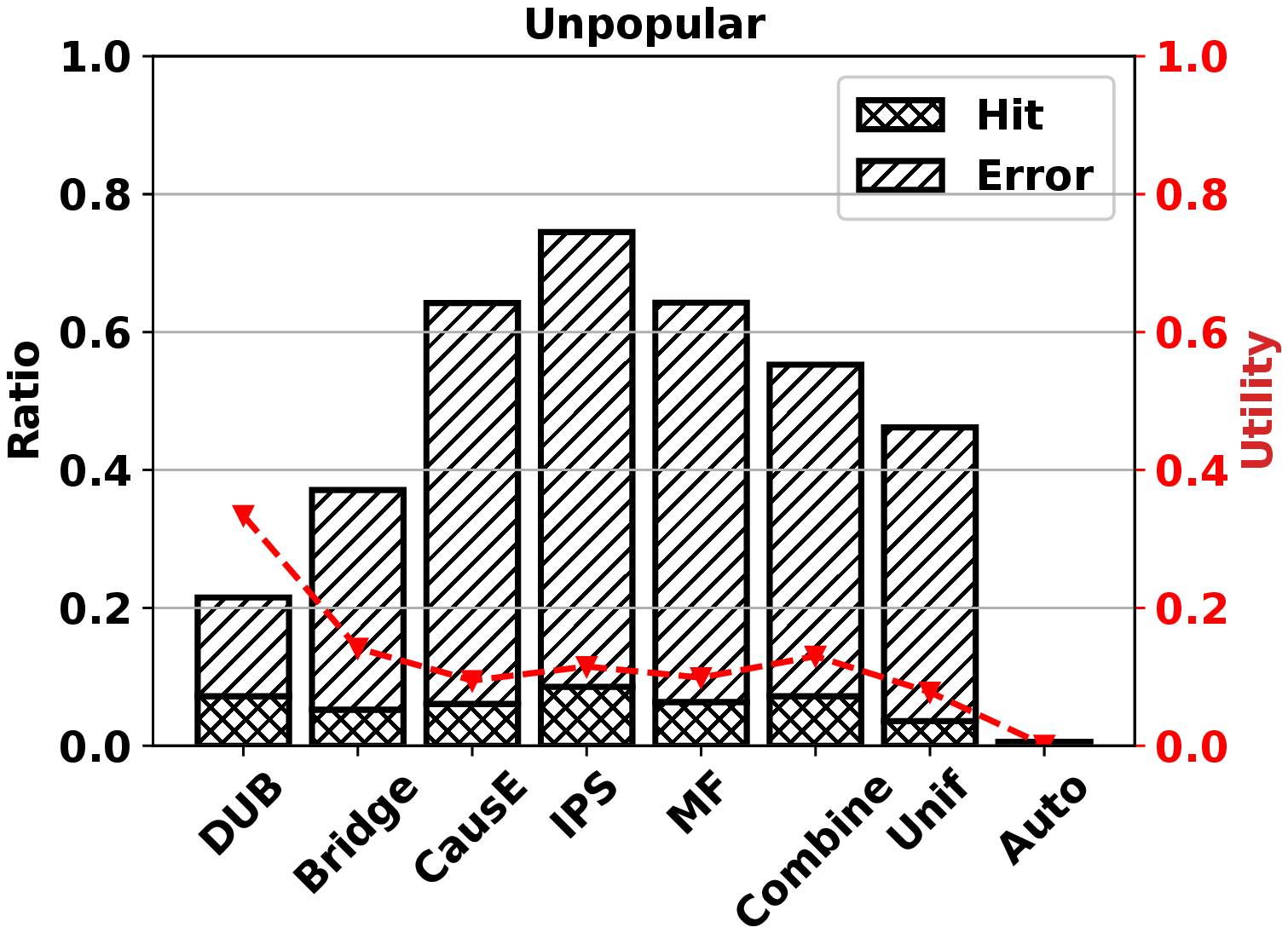}}
\caption{Item distribution and utility of a randomized dataset and different methods with Yahoo! R3.}
\label{fig:ratio}
\end{figure}

An interesting question is about the difference between the distributions of the recommendation lists of the proposed method and the baseline methods.
To answer this question, we show in Figure~\ref{fig:ratio} the item distribution of the recommendation lists generated by different methods, where popular items are the 20\% most frequent items in the training set, and the rest are unpopular items.
Figure~\ref{fig:ratio_random} is the distribution of a randomized dataset, from which we can find that although the probability of popular and unpopular items being recommended is even (e.g., popular items account for 20\% of the total items, and the probability of being recommended also accounts for 20\%), the utility (i.e., the probability of hit divided by the probability of being recommended) brought by popular items is higher. This means that a practical ideal recommendation strategy may not excessively pursue a balance between popular and unpopular items. 
Note that for the brevity of the legend in the figure, we use the abbreviation Auto to refer to the baseline AutoDebias.

Combining Figure~\ref{fig:ratio_pos} and Figure~\ref{fig:ratio_unpos}, we can observe: 
1) MF, IPS and CausE tend to capture the recommendation patterns of popular and  unpopular items similar to Figure~\ref{fig:ratio_random}, but unreasonably displaying too many unpopular items may not bring much benefit, and will even cause user distrust. 
2) AutoDebias can capture the utility information of popular items, but it tends to over-expose the popular items, which may also hurt the user experience.
Note that our results differ somewhat from those in~\cite{chen2021autodebias}. 
As described in Sec.~\ref{subsubsec:datasets}, during data processing, we set the labels of positive and negative feedback to 1 and 0, respectively, to be compatible with the prediction layers with a sigmoid activation.
However, the labels for positive and negative feedback in~\cite{chen2021autodebias} are set to 1 and -1, respectively.
3) Our DUB keeps recommending popular items with high utility, and carefully displays the unpopular items with a higher hit rate and achieves the highest utility among unpopular items, i.e., the DUB can more effectively weigh the use of information between a randomized dataset and a non-randomized dataset. 

\subsection{RQ4: Analysis Results of Key Factors}\label{exp:results_factors}
\begin{figure}
\centering
    \subfigure[]{
    \label{fig:pn_mf}
    \includegraphics[width=0.48\columnwidth]{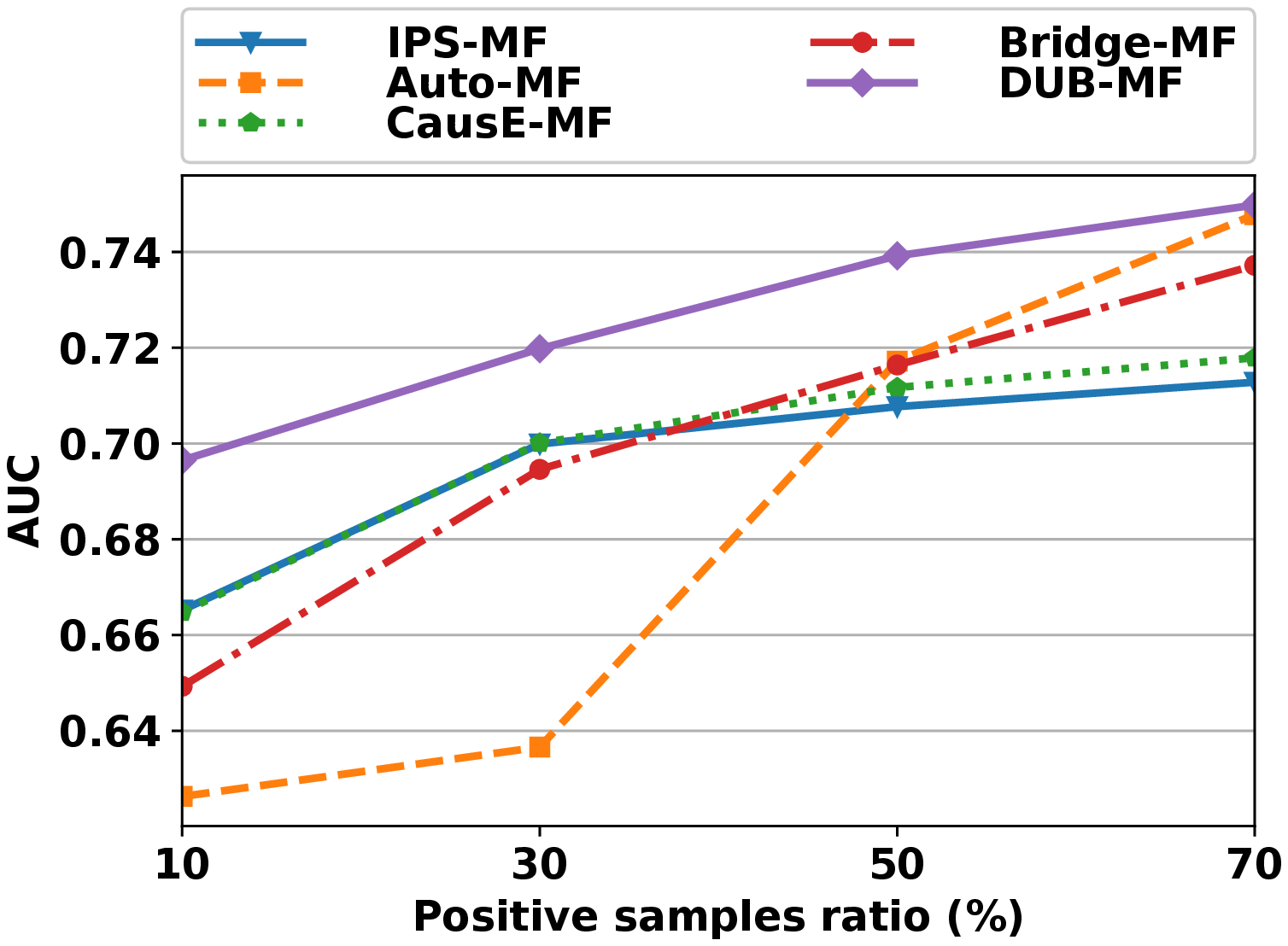}}
    \subfigure[]{
    \label{fig:pn_ncf}
    \includegraphics[width=0.48\columnwidth]{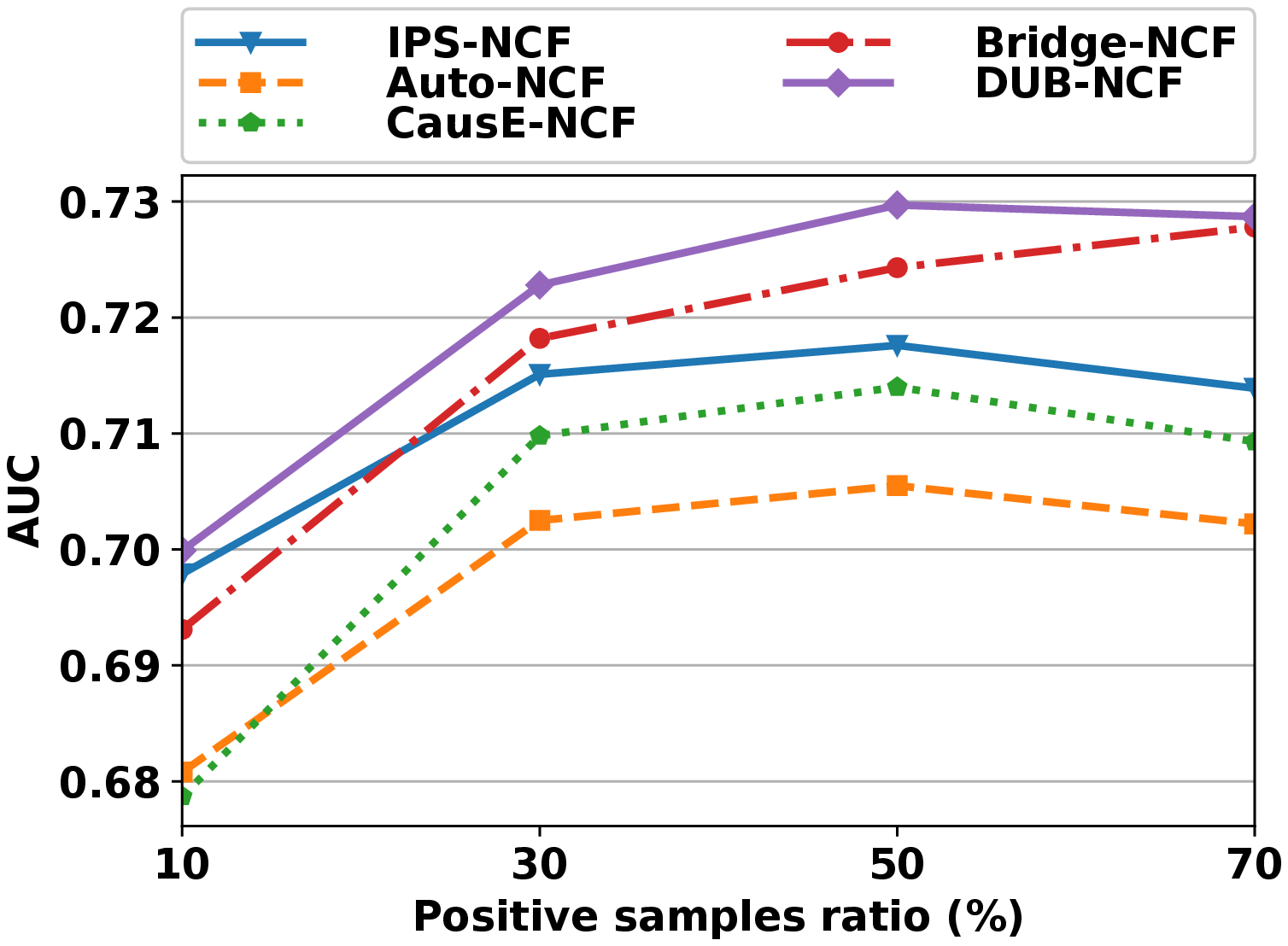}}
    \subfigure[]{
    \label{fig:scale_mf}
    \includegraphics[width=0.48\columnwidth]{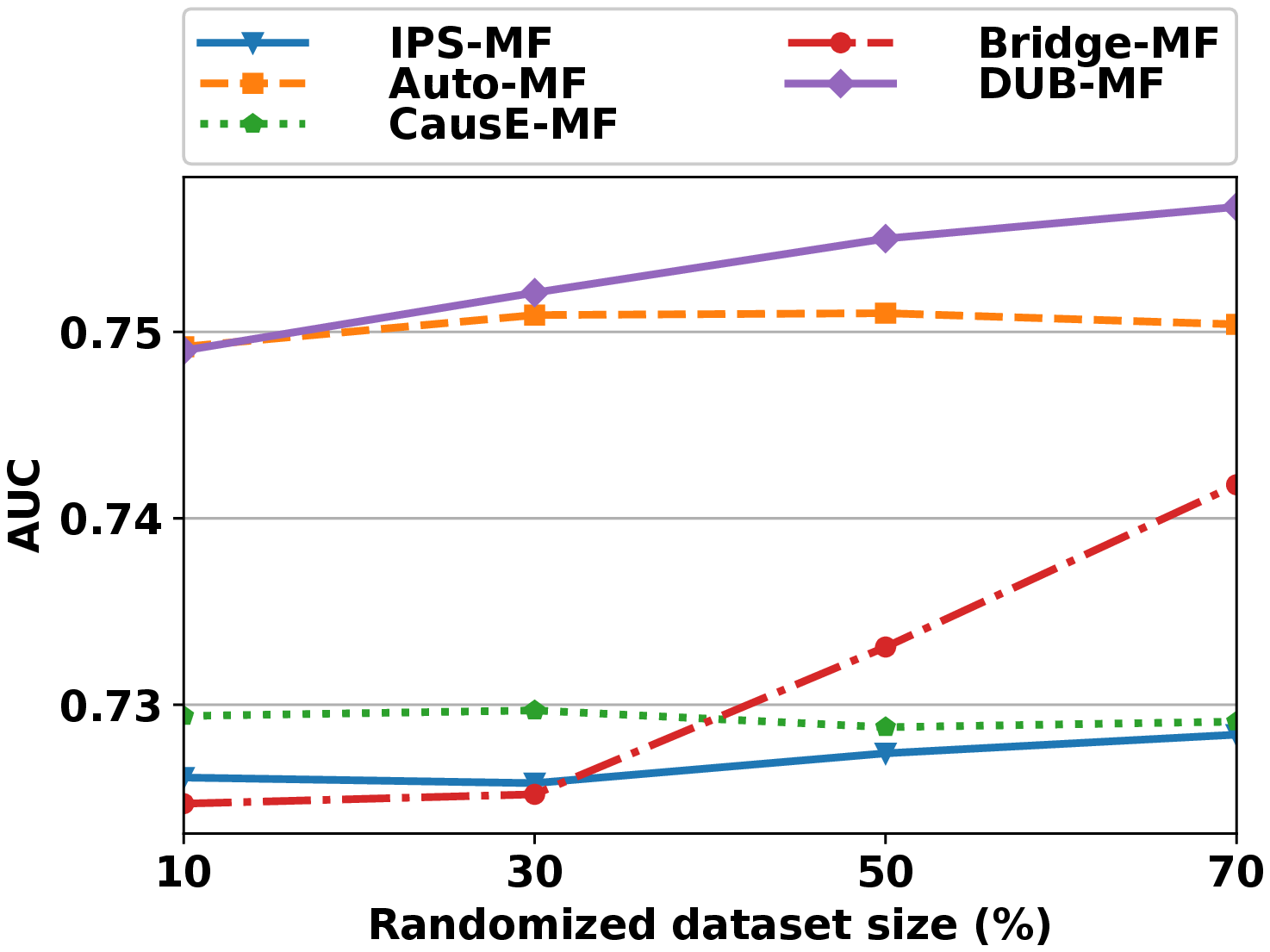}}
    \subfigure[]{
    \label{fig:scale_ncf}
    \includegraphics[width=0.48\columnwidth]{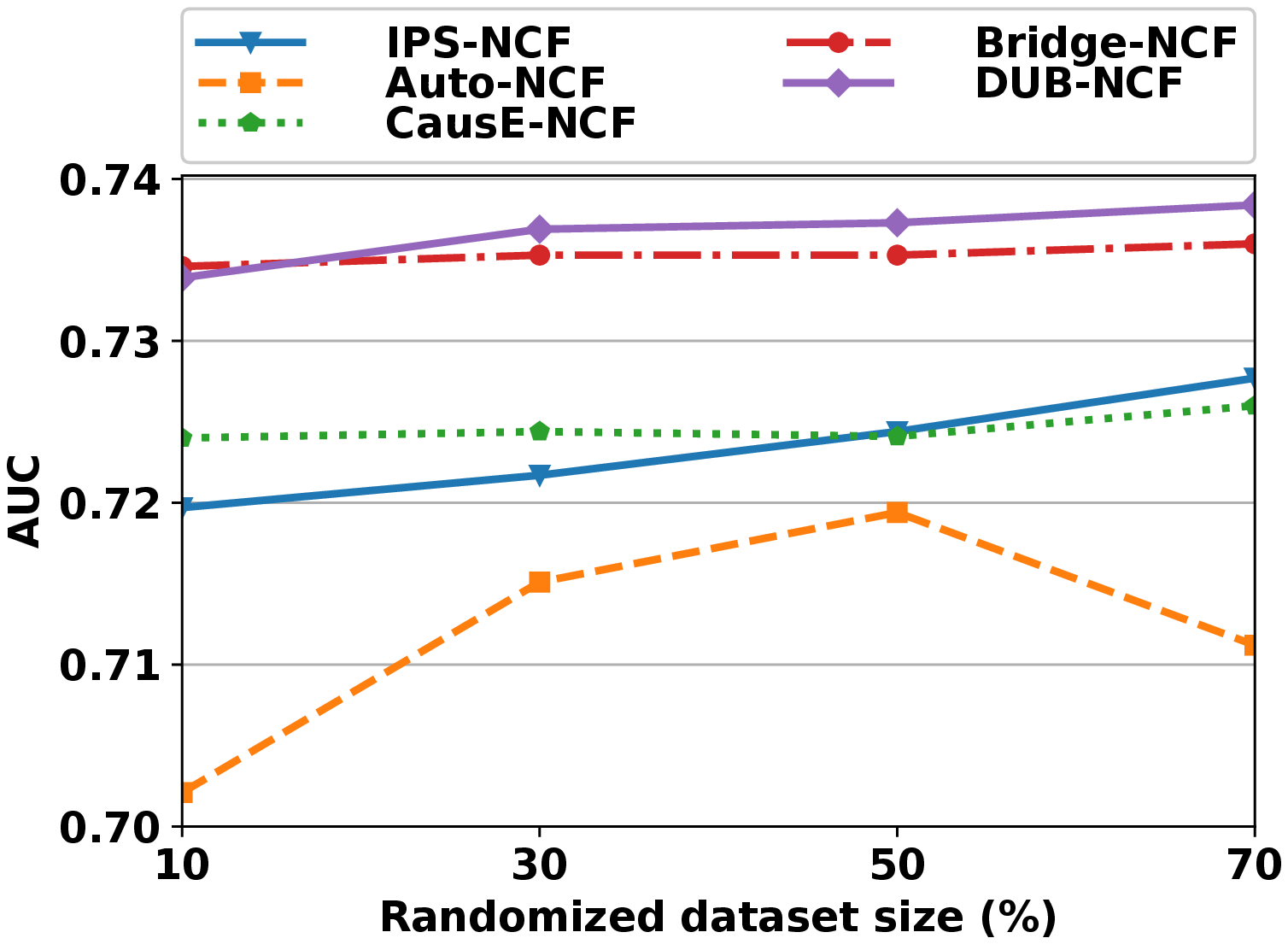}}
\caption{The analysis results of the key factors on Yahoo! R3, where (a) and (b) are considering that $S_c$ has different positive sample ratios, and (c) and (d) are considering that $S_t$ has different data sizes.}
\label{fig:pn}
\end{figure}

We further analyze some key factors that may affect the performance of the methods. 
The first key factor is the difference in the ratio of positive and negative samples between $S_c$ and $S_t$. 
When $S_c$ and $S_t$ are too different, the difficulty of training the model will greatly increase. However, when $S_c$ and $S_t$ are too close, the assimilation will seriously damage the guiding role of $S_t$.
In the experiments, we fix the size of a subset sampled from $S_c$ as 135,000 to include as many positive samples as possible. 
Then we control this subset to contain a certain proportion of positive samples, i.e., we randomly sample $135000*ratio$ positive samples and $135000*(1-ratio)$ negative samples from $S_c$. We set this ratio to 10\%, 30\%, 50\% and 70\%, respectively.
Note that when 10\% is taken, the distribution of this subset is closest to that of $S_t$. 
From Figure~\ref{fig:pn_mf} and Figure~\ref{fig:pn_ncf}, we can find that our DUB consistently outperforms all the baselines in all cases. 

The second key factor that may affect the performance of the model is the size of $S_t$. 
As described in Sec~\ref{subsec:notation}, the scale and scope of $S_t$ is much smaller than that of $S_c$. When the number of $S_t$ is smaller than a certain value, it can hardly guide $S_c$. By observing the performance trend of the model under different sizes of $S_t$, we can have a preliminary understanding of this lower bound.
In the experiments, we keep the same data settings as the previous experiments, except that $S_t$ is randomly sampled according to a certain proportion to obtain a subset. 
We set this ratio to 10\%, 30\%, 50\% and 70\%, respectively.
From Figure~\ref{fig:scale_mf} and Figure~\ref{fig:scale_ncf}, we can find that our DUB is also stable and accurate in all cases.

\subsection{RQ5: Comparison Results of General Evaluation}\label{exp:results_general}
\begin{figure}
\centering
    \subfigure[]{
    \label{fig:4a}
    \includegraphics[width=0.48\columnwidth]{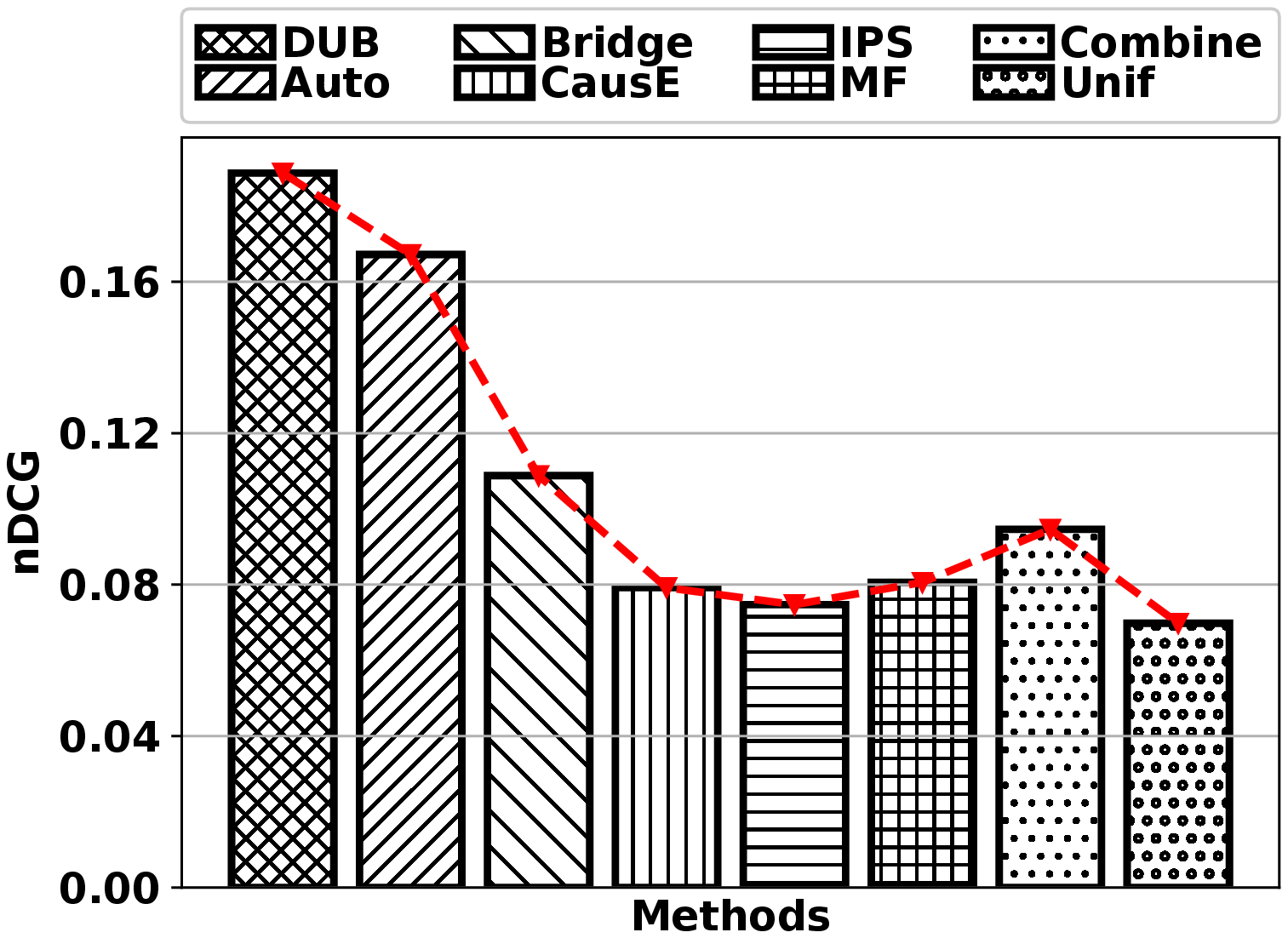}}
    \subfigure[]{
    \label{fig:4b}
    \includegraphics[width=0.48\columnwidth]{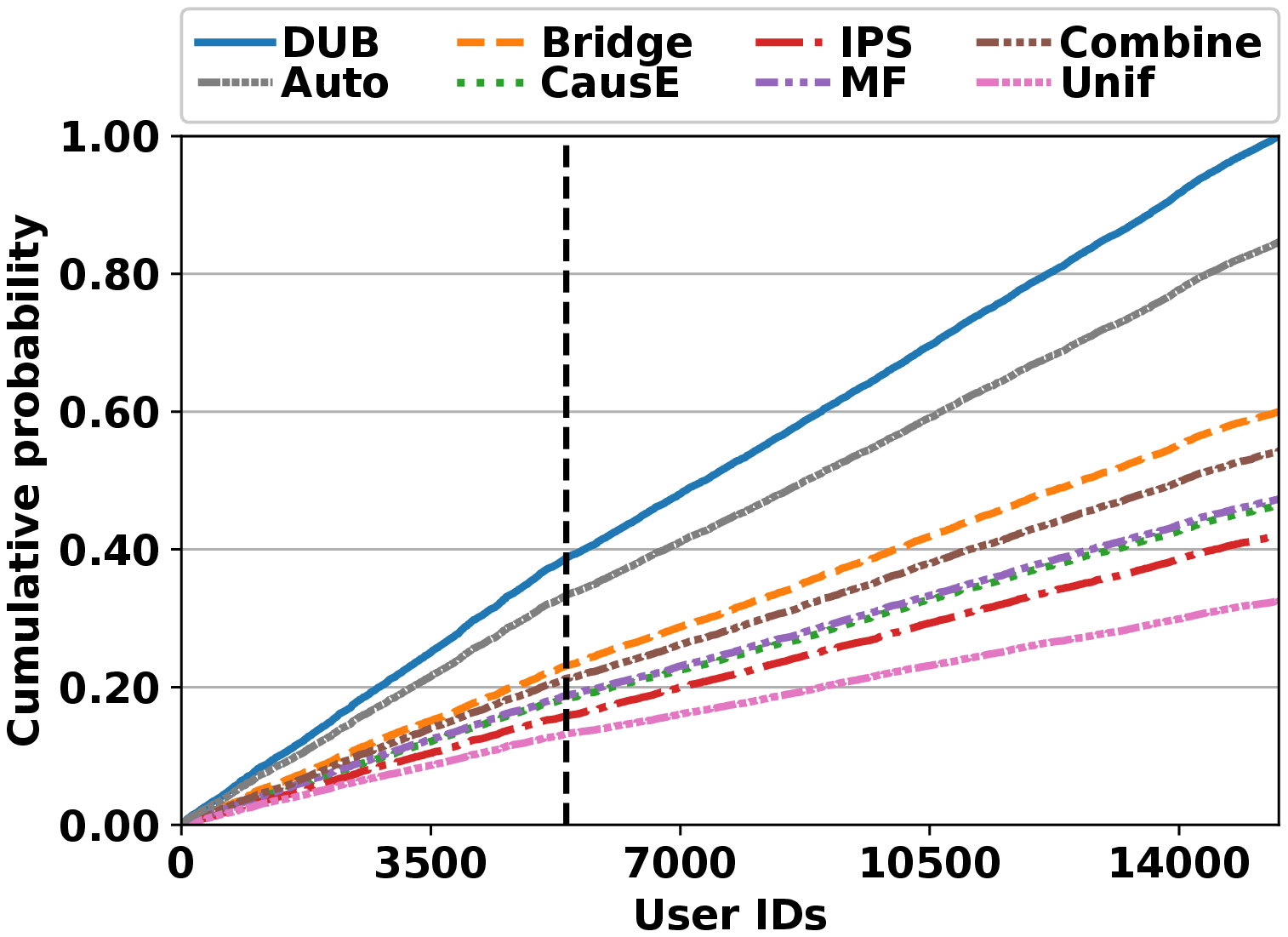}}
\caption{(a) Comparison results in general evaluation. (b) The cumulative hit probability of different methods at the user level. Note that we use Yahoo! R3 in this study.}
\label{fig:4}
\end{figure}

Although using an unbiased data for verification and evaluation is a promising choice, it also has some limitations because it may not cover all the users and items. 
Moreover, we are also interested in the performance of the proposed method and baselines in general evaluation with biased but high coverage, i.e., both validation and testing use the non-uniform data.
In the experiments, we randomly divide $S_c$ according to the proportion of $5:2:3$ to obtain a training set, a validation set and a test set. 
$S_t$ is still used as the unbiased training set. 
We use the same settings in Sec~\ref{subsubsec:imp} to search the best values, except that the reference metric becomes nDCG, because nDCG is one of the most adopted metrics in general evaluation.
We can see from Figure~\ref{fig:4a} that our DUB and AutoDebias have a significant improvement over the other baselines.
This is reasonable because their ability to capture the utility of popular items (as shown in Figure~\ref{fig:ratio}) can play a greater role in general evaluation.
We show in Figure~\ref{fig:4b} the cumulative hit probability of different methods at the user level (i.e., the sum of the hit probabilities of the first $x$ users), and find that introducing $S_t$ in general evaluation is beneficial to better learn the corresponding preferences of the users involved in $S_t$ (i.e., the first 5400 users).

\section{Conclusions and Future Work}\label{sec:feature}
In this paper, we propose a new debiased perspective based on directly optimizing the upper bound of an ideal objective function to facilitate the introduction of some theoretical insights and a more sufficient solution to the system-induced biases.
We first formulate a new unbiased ideal loss function  to more fully reduce the data bias when a small randomized dataset is available, and then give some theoretical insights about its upper bound. 
Moreover, we point out that most existing methods can be regarded as an insufficient optimization of the upper bound. 
As a response, we propose a novel method, i.e., debiasing approximate upper bound with a randomized dataset (DUB), for a more sufficient optimization of the upper bound. 
Finally, we conduct extensive empirical studies to show the effectiveness of the proposed method and explore the impact of some key factors that may affect the performance.

For future works, we will work on obtaining different upper bounds of the unbiased ideal loss function in different ways and comparatively evaluate them.
We also plan to gain more theoretical insights on other ways of using a randomized dataset in debiased recommendation. 
In addition, we are also interested in exploring new techniques for debiased recommendation with only one single non-randomized dataset or multiple non-randomized datasets. 

\begin{acks}
We thank the anonymous reviewers for their expert and constructive comments and suggestions, and the support of National Natural Science Foundation of China Nos. 61836005, 62272315 and 62172283. 
\end{acks}

\bibliographystyle{ACM-Reference-Format}
\bibliography{sample-base}

\appendix

\end{document}